\RequirePackage{fix-cm}
\documentclass[twocolumn]{svjour3}          
\smartqed  
\usepackage{graphicx}
\usepackage{natbib}
\usepackage{amsmath,amsfonts}
\usepackage{booktabs}
\usepackage{wrapfig}
\usepackage{multirow}
\usepackage{graphicx}
\usepackage{subcaption}
\usepackage{url}

\usepackage{xspace}
\def\onedot{.\@\xspace}
\usepackage{mathtools}

\usepackage{enumitem}
\usepackage{multirow}
\usepackage{wrapfig}
\usepackage{colortbl}
\usepackage[normalem]{ulem}
\usepackage{comment}
\usepackage{microtype}
\usepackage[citecolor=black,colorlinks]{hyperref}
\renewcommand{\paragraph}[1]{\vspace{0.5mm}\noindent\textbf{#1.}\,\,}
\usepackage{pifont}

\usepackage{enumitem}
\usepackage{multirow}
\usepackage{colortbl}
\usepackage{microtype}
\usepackage[hang,flushmargin]{footmisc}

\usepackage{amsfonts}

\setlist[itemize]{align=parleft,left=0pt,topsep=1mm,itemsep=0mm,parsep=1mm}

\definecolor{azure(colorwheel)}{rgb}{0.0, 0.5, 1.0}
\definecolor{R5}{rgb}{0.0, 0.7, 0.1}
\definecolor{yw}{rgb}{0.01176, 0.5490, 0.5490}
\definecolor{R123}{rgb}{0.36, 0.54, 0.66}
\definecolor{R1234}{rgb}{0.7, 0.75, 0.71}
\definecolor{applegreen}{rgb}{0.55, 0.71, 0.0}
\definecolor{R132}{rgb}{0.0, 0.0, 1.0}
\definecolor{hs}{rgb}{0.662, 0.482, 0.960}
\definecolor{postechred}{rgb}{0.784, 0.003, 0.313}
\definecolor{gu}{rgb}{0.5460, 0.1755, 0.2766}
\definecolor{ballblue}{rgb}{0.13, 0.67, 0.8}
\definecolor{cornellred}{rgb}{0.7, 0.11, 0.11}
\definecolor{darkcyan}{rgb}{0.0, 0.55, 0.55}
\definecolor{CuGray}{gray}{0.9}
\definecolor{airforceblue}{rgb}{0.36, 0.54, 0.66}
\definecolor{rev}{rgb}{0.784, 0.003, 0.313}
\definecolor{pink}{cmyk}{0, 0.7808, 0.4429, 0.1412}
\definecolor{amethyst}{rgb}{0.6, 0.4, 0.8}
\definecolor{black}{rgb}{0.0, 0.0, 0.0}
\definecolor{tb3_yellow}{rgb}{0.996, 1.0, 0.6}
\definecolor{R123}{rgb}{0.980, 0.8, 0.604}
\definecolor{R512}{rgb}{0.972, 0.6, 0.6}
\definecolor{dimgray}{rgb}{0.41, 0.41, 0.41}
\definecolor{R3}{rgb}{0.8, 0.25, 0.33}
\definecolor{bleudefrance}{rgb}{0.19, 0.55, 0.91}
\definecolor{R6}{rgb}{0.265, 0.445, 0.765}
\definecolor{blue(ryb)}{rgb}{0.01, 0.28, 1.0}
\definecolor{R4}{rgb}{1.0, 0.49, 0.0}
\definecolor{Gray}{gray}{0.88}
\definecolor{green(ncs)}{rgb}{0.0, 0.62, 0.42}
\definecolor{brightpink}{rgb}{1.0, 0.0, 0.5}
\definecolor{alizarin}{rgb}{0.82, 0.1, 0.26}

\definecolor{kellygreen}{rgb}{0.3, 0.73, 0.09}

\newcolumntype{g}{>{\columncolor{CuGray}}c}
\newcolumntype{z}{>{\columncolor{CuGray}}l}

\renewcommand{\paragraph}[1]{\vspace{1mm}\noindent\textbf{#1.}\,\,}
\newcommand{\red}[1]{\textcolor{alizarin}{#1}}

\newcommand{\greencap}[1]{\textcolor{kellygreen}{#1}}

\usepackage{xspace}

\makeatletter
\def\@fnsymbol#1{\ensuremath{\ifcase#1\or *\or \dagger\or \ddagger\or
   \mathsection\or \mathparagraph\or \|\or **\or \dagger\dagger
   \or \ddagger\ddagger \else\@ctrerr\fi}}
\makeatother

\def\onedot{.\@\xspace}
\def\eg{\emph{e.g}\onedot} 
\def\ie{\emph{i.e}\onedot}

\newcommand{\FPRF}{FPRF}

\newcommand{\ours}{FPGS}

\newcommand{\Sref}[1]{Sec.~\ref{#1}}
\newcommand{\Eref}[1]{Eq.~(\ref{#1})}
\newcommand{\Fref}[1]{Fig.~\ref{#1}}
\newcommand{\Tref}[1]{Table~\ref{#1}}

\newcommand{\bmu}{\mbox{\boldmath $\mu$}}

\newcommand{\bsigma}{\mbox{\boldmath $\sigma$}}

\newcommand{\be}{\begin{eqnarray}}
\newcommand{\ee}{\end{eqnarray}}
\newcommand{\bee}{\begin{eqnarray*}}
\newcommand{\eee}{\end{eqnarray*}}

\newcommand{\matrixb}{\left[ \begin{array}}
\newcommand{\matrixe}{\end{array} \right]}

\newcommand{\cmark}{\ding{51}}%
\newcommand{\xmark}{\ding{55}}%

\begin{document}

\title{FPGS: Feed-Forward Semantic-aware Photorealistic Style Transfer of Large-Scale Gaussian Splatting

}


\author{GeonU Kim         \and
        Kim Youwang        \and
        Lee Hyoseok        \and
        Tae-Hyun Oh       
}


\institute{GeonU Kim \at
              School of Artificial Intelligence, POSTECH, Pohang, Republic of Korea \\
              \email{gukim@postech.ac.kr}
           \and
            Kim Youwang \at
              Department of Electrical Engineering, POSTECH, Pohang, Republic of Korea \\
              \email{youwang.kim@postech.ac.kr}
            \and
           Lee Hyoseok \at
              Department of Electrical Engineering, POSTECH, Pohang, Republic of Korea \\
              \email{hyos99@postech.ac.kr}
           \and
           Tae-Hyun Oh \at
              School of Computing, KAIST, Daejeon, Republic of Korea \\
              \email{taehyun.oh@kaist.ac.kr}
}

\date{}

\maketitle

\begin{abstract}
We present \ours, a feed-forward photorealistic style transfer method of large-scale radiance fields represented by Gaussian Splatting.
\ours\, stylizes large-scale 3D scenes with 
arbitrary, multiple style reference images without additional optimization while preserving multi-view consistency and real-time rendering speed of 3D Gaussians.
Prior arts required tedious per-style optimization or time-consuming per-scene training stage and were limited to small-scale 
3D scenes. 
\ours~efficiently stylizes large-scale 3D scenes by introducing a style-decomposed 3D feature field, which inherits AdaIN's feed-forward stylization machinery, supporting arbitrary style reference images. 
Furthermore, \ours~supports multi-reference stylization with the semantic correspondence matching and local AdaIN, which adds diverse user control for 3D scene styles.
\ours~also preserves multi-view consistency by applying semantic matching and style transfer processes directly onto queried features in 3D space.
In experiments, we demonstrate that \ours~achieves favorable photorealistic quality scene stylization for large-scale static and dynamic 3D scenes with diverse reference images.  
Project page: \url{https://kim-geonu.github.io/FPGS/}

\end{abstract}

\section{Introduction}
\label{sec:intro}
Large-scale 3D scene reconstruction is a long-standing problem in computer vision and graphics~\citep{fruh2004automated,snavely2006photo,pollefeys2008detailed, agarwal2009rome, zhu2018very,tancik2022blocknerf}, which aims to build realistic 3D virtual scenes from measurements, \eg, a set of images.
Recently, Neural Radiance Fields~\citep{mildenhall2020nerf, barron2021mip, fridovich2022plenoxels, jun2022hdr, muller2022instant, fridovich2023kplanes} and 3D Gaussian Splatting~\citep{kerbl3Dgaussians} 
have enabled photorealistic reconstruction of 3D scenes by modeling the scenes as radiance fields.
Also, their large-scale extensions~\citep{tancik2022blocknerf, turki2022mega,zhenxing2022switch,  kerbl2024hierarchical,lin2024vastgaussian, liu2024citygaussian} 
have shown remarkable progress in modeling coherent outdoor 3D scenes, suggesting promising future directions, \eg, VR/AR applications.
In this work, we take a step further and address a task, photorealistic style transfer (PST) of large-scale 3D scenes, which we call it 3D scene PST.

The 3D scene PST task aims to 
transfer the visual styles of style reference images onto a large-scale 3D scene represented by a \emph{radiance field}. 
With this objective, the resulting stylized output is expected to be photorealistic and preserve the geometric structure of the original scene.
The large-scale 3D scene PST has various applications,
where it can enrich virtual 3D spaces of XR applications and realistically augment existing real-world autonomous driving datasets~\citep{geiger2012kitti,cordts2016cityscapes}.

Recent studies~\citep{chen2024upst, zhang2023transforming, li2024instant} 
have developed PST methods for 3D scenes represented by neural radiance fields and shown plausible visual qualities.
However, all of them require additional time-consuming training stages to enable feed-forward style transfer~\citep{chen2024upst}
or tedious per-style optimization steps to apply even just a single style reference to the scene~\citep{zhang2023transforming,li2024instant}.
More importantly, due to the aforementioned drawbacks, the previous 3D scene PST methods do not scale to large 3D scenes.
In addition, their neural radiance field representation limits the rendering speed.

\begin{table}[t]
    \setlength\tabcolsep{1.3pt}
        \renewcommand{\arraystretch}{1.2}
    \small
    \caption{Our {FPGS} provides a photorealistic, feed-forward style transfer of 3D Gaussians, while supporting multiple style reference images in the stylization stage. FPGS also achieves real-time rendering, different from competing NeRF-based methods.}
    \begin{tabular}{l cccc}
        \toprule
        & \small{UPST-NeRF} &  \small{LipRF} & \small{StyleGaussian} &  \small{\textbf{FPGS}}\\
        \cmidrule{1-5}
        Photorealistic & \greencap{\cmark} & \greencap{\cmark} & \red{\xmark}&  \greencap{\cmark}\\
        Feed-forward & \greencap{\cmark} & \red{\xmark}& \greencap{\cmark}&  \greencap{\cmark}\\
        Multi-reference & \red{\xmark} & - & \red{\xmark}& \greencap{\cmark}\\
        Real-time render & \red{\xmark} & \red{\xmark}& \greencap{\cmark}& \greencap{\cmark}\\
        \bottomrule
    \end{tabular}

    \label{tab:property}
\end{table}

       

This motivates us to develop an efficient PST method for large-scale 3D scenes that 1) can stylize the whole 3D scene \emph{without} additional time-consuming training or per-style optimization, 2) can render in real-time, and 3) can support diverse multiple reference images.
In this work, we propose a feed-forward PST method of large-scale 3D scenes represented by 3D Gaussian Splattings (FPGS).
To implement a feed-forward 3D PST method, we employ the adaptive instance normalization (AdaIN)~\citep{huang2017adain} layer before the rasterization process to dress up the decoded colors with the styles.
Specifically, we propose a photorealistic \emph{stylizable radiance field} consisting of a scene \emph{content} field and a scene \emph{semantic} field.
Given a large set of photos of the target scene, we first train a scene \emph{content field} to embed the scene geometry and content features that can be later decoded to a large-scale radiance field of arbitrary styles.
The scene \emph{semantic field} is trained together to aid to match proper local styles to the local scene.
With the obtained scene content field, FPGS stylizes the whole 3D scene via the feed-forward AdaIN, which manipulates the scene content field with the 
style reference images' feature statistics in a feed-forward manner without any nuisance optimization. 
After the feed-forward style transfer, we can render the stylized scene in real-time, which is not achieved by prior methods~\citep{chen2024upst, zhang2023transforming, li2024instant} based on neural radiance fields.

In addition to efficiency, stylizing a large-scale scene requires dealing with diverse objects and contents that are not covered with a single reference image. 
Also, it is challenging to identify a single reference that can effectively encompass the entire semantics of a large-scale scene. 
Thus, extending the existing single reference-based methods~\citep{chen2024upst, liu2023stylerf, li2024instant} is not straightforward due to diverse contents in the large-scale scene.
To overcome this challenge, we propose a style dictionary module and style attention for efficiently retrieving the style matches of each local part of the 3D scene from a given set of diverse style references.
The proposed style dictionary consists of pairs of a local semantic code and a local style code extracted from the style references.
To form a compact style dictionary, 
we exploit the clustering of styles and semantics that notably reduces the computational complexity of the style retrieval.
Using a style dictionary, we find semantic correspondences between the local semantic codes and the style semantic field.

This work is an extension of our conference version, FPRF~\citep{kim2024fprf}, which focuses on stylizing static 3D scenes represented by neural radiance fields. 
We notably extend its representation, process, computational speed, and applications.
We achieve real-time rendering of the stylized scene by two extended features: 1) representing scene with 3D Gaussians instead of neural radiance field and 2) redesigning the stylization transfer process.
Specifically, while FPRF requires a stylization process for each new view rendering due to its implicit representation, FPGS stylizes the whole scene at once before rendering, leveraging the explicit nature of 3D Gaussians.
This decoupling of stylization and rendering process enables real-time rendering after a feed-forward scene style transfer. 
We also propose a compact per-RGB -to- scene content feature encoder for stylization, called
MLP VGGNet, which does not require any training or feature distillation, and an iterative style transfer method which enhances local style transfer quality.
We further extend the target domain to 4D scenes including large-scale urban scenes, which has not been explored in previous 3D PST methods~\citep{chen2024upst, zhang2023transforming, li2024instant,kim2024fprf}. 
In this work, we introduce both FPRF and FPGS to clearly contrast the benefits of FPGS.

Our experiments demonstrate that
our FPGS obtains superior stylization qualities on both large/small-scale scenes with multi-view consistent and semantically matched style transfer.
We also show the exceptional efficiency of FPGS in terms of training and rendering time.
Furthermore, we demonstrate versatility of our method with various applications including 4D scene stylization, multi-reference stylization, and scribble-based stylization, which are not supported by other prior methods.
Our main contributions are summarized as follows:
\begin{itemize}
    \item We propose a stylizable radiance field represented by 3D Gaussians where we can perform photorealistic style transfer in a feed-forward manner and achieve real-time rendering.
    \item We propose the style dictionary and its style attention for style retrieval, which allows us to deal with multiple style references via semantic correspondence matching.
    \item 
    To the best of our knowledge, 
    our work is the first multi-reference based 3D/4D PST without any per-style optimization, which is scalable for large-scale scenes. 
\end{itemize}

\section{Related Work}
\label{sec:related}
\noindent{Our task relates to large-scale scene reconstruction and photorealistic style transfer for large-scale 3D scenes.}

\paragraph{Large-scale 3D scene reconstruction}
Realistic 3D reconstruction of large-scale scenes
has been considered as an important task, which could be a stepping stone to achieve a comprehensive 3D scene understanding and immersive virtual reality.
Recently, several studies~\citep{tancik2022blocknerf,turki2022mega,zhenxing2022switch, kerbl2024hierarchical,lin2024vastgaussian, liu2024citygaussian} tackled the task based on advances in Neural Radiance Fields~\citep{mildenhall2020nerf} and 3D Gaussian Splatting~\citep{kerbl3Dgaussians}, which show remarkable reconstruction qualities for large-scale 3D scenes.
Prior arts mainly focused on decomposing the large-scale scene into smaller parts, \ie, divide-and-conquer.
NeRF-based methods~\citep{tancik2022blocknerf, turki2022mega, zhenxing2022switch} divide large-scale scenes into multiple regions and optimize a neural radiance field for each region. 
Similarly, Gaussian-based methods~\citep{lin2024vastgaussian, kerbl2024hierarchical, liu2024citygaussian} also reconstruct a large-scale scene in a divide-and-conquer manner and achieve real-time rendering.
We propose methods for the two representations, FPRF is for NeRF-based methods, FPGS is for Gaussian-based methods.

The large-scale scene comprising multiple radiance fields poses a challenge for style transfer in terms of efficiency. 
Stylizing a large-scale scene with 
per-style optimization methods~\citep{zhang2023transforming, li2024instant} is impractical, which requires extensive time to optimize every radiance field in the large-scale scene whenever new style images are given.
We overcome this limitation by transferring style of radiance fields in a feed-forward manner. 
Our models are trained separately with an efficient training stage compared to previous feed-forward methods~\citep{chen2024upst, liu2023stylerf}, and generalized to any scenes and any style reference images. 

\paragraph{Photorealistic style transfer of 3D scenes}
The 3D scene style transfer task aims to stylize 3D scenes according to the style of the given reference image while preserving multi-view consistency. 
Recent works \citep{chiang2022stylizing, huang2022stylizednerf, fan2022unified, nguyen2022snerf, chen2024upst, zhang2022arf, liu2023stylerf, zhang2023transforming, li2024instant} combine 3D neural radiance fields with style transfer methods. 
Among them, UPST-NeRF~\citep{chen2024upst}, Instant-NeRF-Stylization~\citep{li2024instant} and LipRF~\citep{zhang2023transforming} tackled PST on 3D neural radiance fields. 
UPST-NeRF constructs a stylizable 3D scene with a hypernetwork, which is trained on stylized multi-view images of a single scene.
They can stylize a trained scene in a feed-forward manner with arbitrary style. However, the model requires additional time-consuming (${>}10$ hrs.) per-scene optimization after scene reconstruction.
Instant-NeRF-Stylization trains the content and style sub-branches and executes AdaIN~\citep{huang2017adain} to perform 3D style transfer.
LipRF stylizes the reconstructed 3D scene with multi-view stylized images with 2D PST methods, by leveraging Lipshitz network~\citep{virmaux2018lipschitz}.
However, Instant-NeRF-Stylization and LipRF require a time-consuming iterative optimization process for every unseen reference style image.
One of the artistic style transfer methods, StyleRF~\citep{liu2023stylerf} leverages a 3D feature field to stylize 3D scenes with an artistic style.
They achieve feed-forward style transfer with distilled 3D feature field, however, it also requires time-consuming per-scene training stages which spends much more time than reconstruction (${>}5$hrs.) even for a small scene. 
Recently, StyleGaussian~\citep{liu2024stylegaussian} proposes style transfer of radiance fields represented by 3D Gaussians, which achieves multi-view consistent style transfer and real-time rendering. 
However, StyleGaussian focuses on artistic style transfer without considering semantic correspondence between the scene and the reference image, and also requires an additional time-consuming training process ($>$5hrs.).

Our method, FPGS, mitigates aforementioned inefficiency by constructing stylizable 3D radiance field 
represented by 3D Gaussians with a faster training stage spending about 24 mins.
Also, we distinctively focus on photorealistic stylization with the capability of referring to multi-style references, while StyleGaussian focuses only on artistic stylization with a single reference.
\Tref{tab:property} compares the distinctiveness of our method with the prior works.

\section{Preliminary}
\subsection{Neural Radiance Fields (NeRF)}
\noindent
A radiance field $f$ is a continuous 5D function that maps any 3D point $\mathbf{x} \in \mathbb{R}^3$ and viewing direction $\mathbf{d} \in \mathbb{S}^2$ to a volume density $\sigma \in \mathbb{R}^+$ and a color $\mathbf{c} \in \mathbb{R}^3$.
Neural Radiance Fields (NeRF)~\citep{mildenhall2020nerf} use MLP to represent radiance fields, and the scene representation can be rendered and optimized via differentiable volume rendering.
A pixel's color $\hat{\mathbf{C}}(\textbf{r})$ of given ray $\mathbf{r}$ is computed by accumulating the color \(\hat{\mathbf{c}}_i\) and density \(\sigma_i\) of $N_\mathbf{r}$ sampled 3D points
\(\mathbf{x}_i\) along the ray \textbf{r} as:
\begin{equation}
\begin{gathered}
\hat{\mathbf{C}}(\mathbf{r}) = \sum_{i=1}^{N_\mathbf{r}}
{\hat{\mathbf{c}}}_i\alpha^\text{NF}_i\prod^{i-1}_{j=1}(1-\alpha^{\text{NF}}_j)\\
\alpha^{\text{NF}}_j = 1-\text{exp}(-\sigma_i\delta_i),
\label{eq:nerf color volume rendering}
\end{gathered}
\end{equation}
where \(\delta_i\) denotes the distance 
between the sampled point $\mathbf{x}_i$ and the next sample point, \(\alpha^{\text{NF}}_j\) is the absorption by \(\mathbf{x}_i\), 
and \(\prod^{i-1}_{j=1}(1-\alpha^{\text{NF}}_j)\) denotes the transmittance to the point \(\mathbf{x}_i\).
NeRF is optimized by minimizing the rendered pixel values and ground truth pixel values from training images.
Recently, grid-based approaches~\citep{chen2022tensorf, fridovich2023kplanes, muller2022instant, jun2024factorized} utilize efficient grid backbones to accelerate training and rendering speed.
We adopt K-planes~\citep{fridovich2023kplanes} as the neural radiance field representation of \FPRF~due to its efficiency.

\subsection{3D Gaussian Splatting (3DGS)}
\label{sec:3dgs preliminary}
Different from NeRF, 3D Gaussian Splatting (3DGS) \citep{kerbl3Dgaussians} represents a radiance field with explicit 3D Gaussian primitives.
In detail, 3D Gaussians are parameterized as: $\mathcal{G} = \{\textbf{g}_i = (\mathbf{p}_i, \mathbf{\Sigma}_i, \sigma_i, \mathbf{K}_i)\}_{i=1,...,P}$, where $P$ is the number of 3D Gaussians, a mean $\mathbf{p}_i \in \mathbb{R}^3$ denotes the position of a 3D Gaussian $\textbf{g}_i$ and a covariance $\mathbf{\Sigma}_i \in \mathbb{R}^{3\times3}$ denotes the scale and shape of $\textbf{g}_i$.
An opacity $\mathbf{\sigma}_i \in \mathbb{R}^+$ and
spherical harmonics (SH) coefficients $\mathbf{K}_i \in \mathbb{R}^{3\times16}$ specify the geometry and appearance of the scene.
A color $\hat{\mathbf{c}}_i$ of a 3D Gaussian $\hat{\textbf{g}}_i$ can be decomposed into a view-independent base color $\hat{\mathbf{c}}_i^\text{diff}$ and a view-dependent color $\hat{\mathbf{c}}_i^\text{spec}$ 
which are obtained via SH representation as:
\begin{equation}
    \begin{gathered}        
    \hat{\mathbf{c}}_i = \hat{\mathbf{c}}_i^\text{diff} + \hat{\mathbf{c}}_i^\text{spec}, \\
    \hat{\mathbf{c}}_i^\text{diff} = 
    \mathbf{K}_i^{[:,1:2]}C_\text{SH}, ~~
    \hat{\mathbf{c}}_i^\text{spec} = 
    \mathbf{K}_i^{[:,2:17]}\Gamma_\text{SH}(\mathbf{d}).
   \end{gathered}
    \label{eq:spherical harmonics}
\end{equation}
A matrix $\mathbf{K}_i^{[:,n:m]}$ comprises the columns of $\mathbf{K}_i$ from $n$th up to $m$th column, $C_\text{SH}$ is the pre-defined view-independent constant, and $\Gamma_\text{SH}$ is the basis function which generates a basis $\Gamma_\text{SH}(\mathbf{d}) \in \mathbb{R}^{15}$ depending on a view direction $\mathbf{d}$.
Similar to \Eref{eq:nerf color volume rendering}, a pixel's color $\hat{\mathbf{C}}(\mathbf{r})$ 
 of a given ray $\mathbf{r}$ is computed as: 
\begin{equation}
\hat{\mathbf{C}}(\mathbf{r}) = \sum_{\textbf{g}_i \in \mathcal{G}_\mathbf{r}}
{\hat{\mathbf{c}}}_i\alpha^\text{GS}_i\prod^{i-1}_{j=1}(1-\alpha^{\text{GS}}_j),
\label{eq:3dgs color volume rendering}
\end{equation}
where $\mathcal{G}_\mathbf{r}$ is the ordered set of overlapped 3D Gaussians on a ray $\mathbf{r}$ and $\alpha^{\text{GS}}_i$ is obtained by projecting the 3D Gaussian $\textbf{g}_i$ to 2D Gaussian and multiplying the density of the 2D Gaussian with $\sigma_i$.
The 3D Gaussians $\mathcal{G}$ are optimized by minimizing a photometric loss between the rendered images and training images.
3DGS provides a tile-based splatting solution achieving real-time rendering of the scene, which is notably faster than prior neural volume rendering methods. We leverage this property of 3DGS to achieve real-time rendering after the scene stylization.
%

\section{Feed-Forward Photorealistic Style Transfer of Neural Radiance Field - FPRF}\label{sec:FPRF}
\noindent{In this section, we propose FPRF, a feed-forward Photorealistic Style Transfer (PST) method of neural radiance fields, as a reference method of feed-forward stylizable radiance field, FPRF.
Then, we extend to Gaussian-based one, FPGS, in \Sref{sec:FPGS}, which is an advanced PST method for Gaussian splatting.}
We first propose a single-stage training of the stylizable radiance field using AdaIN (\Sref{sec:pst_adain}). 
We further describe the scene semantic field to stylize large-scale scenes with multiple reference images (\Sref{sec:pst_semantic_matching}).

\begin{figure}[t]
    \centering
    \includegraphics[width=\linewidth]{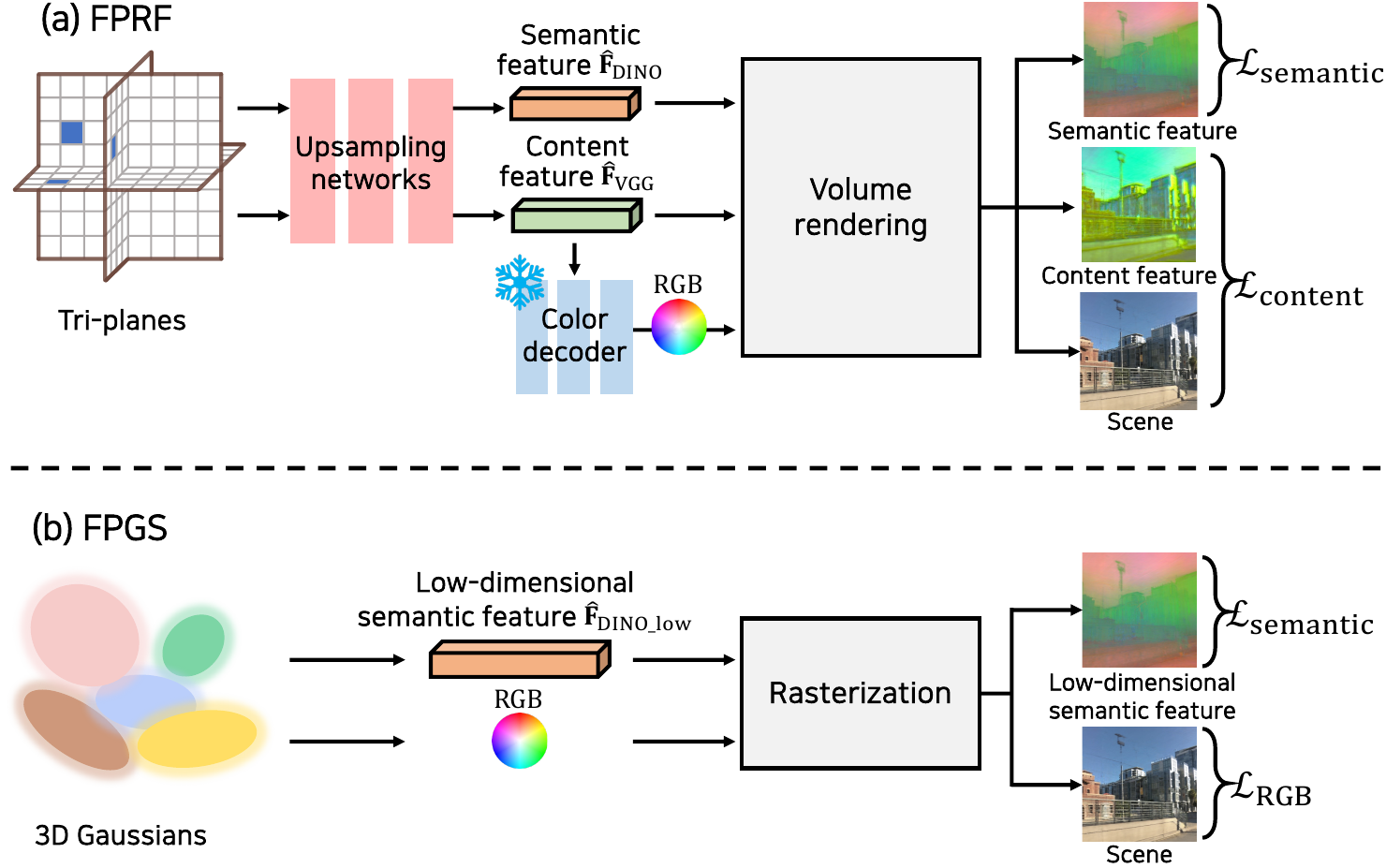}
    \caption{\textbf{Training stage.} Given the scene images, our methods learn the stylizable radiance field.
    \textbf{[Top]} FPRF reconstructs the scene semantic field and the scene content field by distilling high-dimensional semantic and content features extracted from the training images.
    \textbf{[Bottom]} Different from FPRF, FPGS reconstructs the scene content field with RGB loss only and learns the scene semantic field with compressed low-dimensional semantic features.
    }
    \label{fig:pipeline_train}
\end{figure}

\subsection{Large-scale Stylizable Radiance Field}
\label{sec:pst_adain}
\noindent{Our goal is to construct a large-scale radiance field that can be stylized with reference images in a feed-forward manner while achieving photorealistic results after stylization.
}
For the feed-forward style transfer, we employ adaptive instance normalization (AdaIN)~\citep{huang2017adain}, which is an efficient style reference-based style transfer method, where it linearly transforms the content image's feature statistics with a given style image's feature as:
\begin{equation}
    \text{AdaIN}(\mathbf{F}(c), \mathbf{F}(s)) =\bsigma_s\frac{(\mathbf{F}(c) - \bmu_c)}{\bsigma_c} + \bmu_s,
\end{equation}
where \(\mathbf{F}(c)\) and \(\mathbf{F}(s)\) are semantic features extracted from content and style images by a pre-trained feature encoder, \eg, VGGNet~\citep{simonyan2014very}. 
The feature statistics \(\bmu_{*}\) and \(\bsigma_{*}\) over spatial axes are the mean and standard deviation of the extracted features, respectively.
The AdaIN layer is favorable for enabling feed-forward style transfer.
This property is particularly useful when dealing with large-scale 3D scenes and enables a fast and seamless stylization.
To leverage AdaIN's statistic-based style transfer to the 3D scene,
we propose reconstructing a stylizable 3D neural radiance field by distilling the 2D features onto the field, called the stylizable radiance field.

\paragraph{Stylizable radiance field}
To build a multi-view consistent stylizable feature field, 
we apply multi-view bootstrapping~\citep{simon2017hand} to our domain.
We distill high-dimensional features obtained from 2D input images into neural feature fields that models the 3D scene. 
To represent a large-scale 3D scene, we extend 
K-planes~\citep{fridovich2023kplanes} with the block-composition manner~\citep{tancik2022blocknerf}, which is used for embedding volumetric scene geometry, radiance, and semantic features.
Specifically, we design the stylizable radiance field
with two tri-plane grids: scene \emph{content} field and scene \emph{semantic} field. 
The scene semantic field embeds semantic features of a scene, which will be discussed later in \Sref{sec:pst_semantic_matching}.
The scene content field is responsible for embedding 
accurate scene geometry and appearance-related content features.
Given a 3D scene point position $\mathbf{x}{=}[x, y, z]$ and a ray direction vector $\mathbf{d}$ as inputs, our scene content field outputs
the density, and content feature of the query 3D point (see \Fref{fig:pipeline_train}).

The original NeRF computes a pixel's RGB color $\hat{\mathbf{C}}(\textbf{r})$ with \Eref{eq:nerf color volume rendering}.
Correspondingly, we train to render high-dimensional features of 3D points in the scene to a pixel value $\hat{\mathbf{F}}(\mathbf{r})$ by accumulating features \(\hat{\mathbf{f}}_i\) along the ray $\mathbf{r}$:

\begin{equation}
\hat{\mathbf{F}}(\mathbf{r}) = \sum_{i=1}^{N_\mathbf{r}}
{\hat{\mathbf{f}}}_i\alpha^\text{NF}_i\prod^{i-1}_{j=1}(1-\alpha^{\text{NF}}_j),
\label{eq:feature volume rendering}
\end{equation}
where we substitute the high-dimensional feature $\hat{\mathbf{f}}$ for the color $\hat{\mathbf{c}}$ of \Eref{eq:nerf color volume rendering}.
We distill 2D image features to a 3D scene by minimizing the error between the volume-rendered features and the 2D features extracted from input images, which we call feature distillation.
To distill highly-detailed features into the
3D scenes, we use the refined 2D features by Guided Filtering~\citep{he2012guided} before distillation.
We employ upsampling MLP networks to match the dimension of features from the tri-plane to the dimension of distilled high-dimensional features.

\begin{figure*}[t!]
    \centering
    \includegraphics[width=\linewidth]{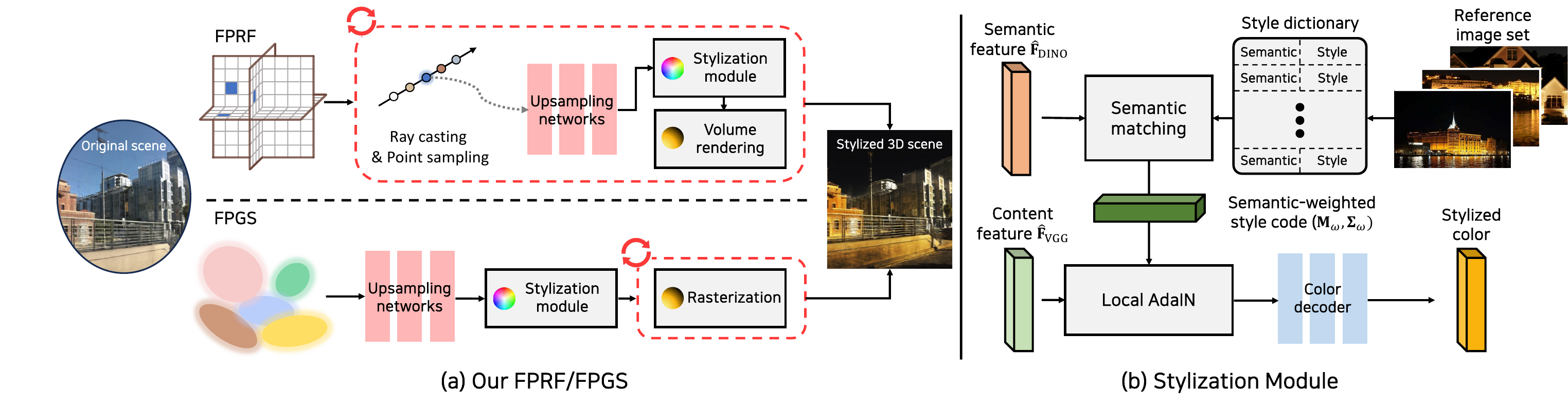} 
    \caption{\textbf{Stylization stage}.
    (a) The stylization module and rendering process of FPRF are intertwined so that the time-consuming stylization process is required for each new view rendering. In contrast, FPGS stylizes whole 3D Gaussians at once before rendering, and the stylized scene can be rendered in real-time.
    (b) Given the optimized stylizable radiance field and the set of arbitrary reference images, we stylize the large-scale 3D scene via our novel semantic-aware local AdaIN. 
    We compose a style dictionary consisting of local semantic codes and local style codes pairs extracted from the clustered reference images. 
    Using semantic features from the stylizable radiance as a query, we find the corresponding local semantic features and retrieve the paired local style codes. 
    Using the retrieved semantic-style code pairs, we perform semantic matching and local AdaIN, then finally render the stylized colors.
    }
    \label{fig:pipeline_inference}
\end{figure*}

\paragraph{Generalizable pre-trained MLP color decoder}
The output of the scene content field requires a separate decoder to decode the stylized feature into the color.
One may na\"ively train such a decoder scene-by-scene, which  suffers from a limited generalization to unseen colors and requires additional training for stylization.
In the following, we present a scene-agnostic and pre-trained MLP color decoder \(D_\text{VGG}\) compatible with AdaIN.
Specifically, \(D_\text{VGG}\) transforms the distilled VGG features into colors.
To perform style transfer in a feed-forward manner, we pre-train \(D_\text{VGG}\) with the diverse set of content images~\citep{lin2014microsoft} and style images~\citep{nichol2016painter}, which enable \(D_\text{VGG}\) to be generalized to arbitrary style reference images.
For that, we employ content loss \(\mathcal{L}_c\) and the style loss \(\mathcal{L}_s\), similar to \citep{huang2017adain}: $\mathcal{L}_{D_\text{VGG}} = \mathcal{L}_c + \lambda_{s}\mathcal{L}_s$.
The main differences with the training process of AdaIN are threefold.
First, we use the MLP architecture instead of the CNN upsampling layer inducing multi-view inconsistency.
Also, we employ features from the shallower layer, \texttt{ReLU2\textunderscore1} of VGGNet, that contain richer color information.
Furthermore, the features from the input content images are upsampled to the pixel resolution for per-pixel decoding.
When training the stylizable radiance field, we fix the pre-trained $D_\text{VGG}$ so that it preserves the knowledge about the distribution of VGG features from diverse images, which induces compatibility with AdaIN.

\paragraph{Training scene content field}
We train the scene content field by optimizing
the tri-plane grid features and MLP.
In detail, for the input 3D point $\mathbf{x}_i$ and the view direction vector $\mathbf{d}$, the grid features of the scene content field are decoded into the 3D point density $\sigma_i$ and the content feature (see the green grid and MLP in \Fref{fig:pipeline_train}).
We further decode the content feature into RGB values using the pre-trained $D_\text{VGG}$.
At the end, we render the content feature \(\hat{\mathbf{F}}_\text{VGG}^\text{NF}(\textbf{r})\) and the scene color \(\hat{\mathbf{C}}(\textbf{r})\) via volume rendering (\Eref{eq:feature volume rendering}).

To perform AdaIN, we guide the content feature map ${\hat{\mathbf{F}}}_\text{VGG}^\text{NF}$ to follow VGG feature distribution by feature distillation.
We distill the ground truth features \(\mathbf{F}_\text{VGG}(\mathbf{I})\) obtained from input images \(\mathcal{I} = \{\mathbf{I}^i\}_{i=1...C}\) 
via pre-trained VGGNet, by minimizing the error between \(\mathbf{F}_\text{VGG}(\mathbf{I})\) and the volume rendered features \({\hat{\mathbf{F}}}_\text{VGG}^\text{NF}(\textbf{r})\).
Note that the ground truth VGG feature maps are upsampled to pixel resolution with guided filtering.
Since the scene content field needs to reconstruct 
the accurate scene geometry and appearance, we compute the photometric loss for the volume rendered color \(\hat{\mathbf{C}}(\mathbf{r})\).
Also, typical regularization losses $\mathcal{L_\text{reg}}$ are employed~\citep{fridovich2023kplanes}.
The total loss function for training the scene content field is as below:
\begin{multline}
    \mathcal{L}_\text{content} = \sum_{\mathbf{r} \in \mathcal{R}} \lVert \hat{\mathbf{F}}_\text{VGG}^\text{NF}(\mathbf{r}) - \mathbf{F}_\text{VGG}(\mathbf{I}, \mathbf{r})\rVert^2_2 \\ + 
    \lambda_{\text{RGB}}\sum_{\mathbf{r} \in \mathcal{R}} \lVert \hat{\mathbf{C}}(\mathbf{r}) - \mathbf{C}(\mathbf{I}, \mathbf{r})\rVert^2_2  + 
    \lambda_\text{reg}\mathcal{L}_\text{reg},
\label{eq:loss_color}
\end{multline}
where \(\mathcal{R}\) is the set of sampled rays in each training batch, and \(\mathbf{F}_\text{VGG}(\mathbf{I}, \mathbf{r})\) and \(\mathbf{C}(\mathbf{I}, \mathbf{r})\) denote ground truth VGG features and RGB values of the pixels correspond to the ray \(\mathbf{r}\in\mathcal{R}\). 

Note that we keep \(D_\text{VGG}\) frozen after its pre-train stage, \ie, \(D_\text{VGG}\) is fixed during the training stage of the scene content field. This differs from StyleRF~\citep{liu2023stylerf}, which needs to fine-tune a CNN-based decoder for each scene.

\paragraph{Feed-forward stylization using the scene content field}
After training the scene content field,
we can perform PST with an arbitrary style image in the stylization stage.
In other words, we train the scene content field once and perform PST on a trained radiance field in a feed-forwards manner, \ie, without per-style or per-scene optimization.

Given a reference image 
\(\mathbf{I}_s\),
we start with extracting the VGG feature \(\mathbf{F}_\text{VGG}(\mathbf{I}_s)\).
Then we stylize 3D content features \(\hat{\mathbf{F}}_\text{VGG}^\text{NF}\) as \( \bsigma_s\frac{(\hat{\mathbf{F}}_\text{VGG}^\text{NF} - \bmu_c)}{\bsigma_c} {+} \bmu_s\), where \(\bmu_s\) and \(\bsigma_s\) are the mean and standard deviation of \(\mathbf{F}_\text{VGG}(\mathbf{I}_s)\).
By decoding the stylized content features to RGB values using the pre-trained color decoder \(D_\text{VGG}\), 
we can render a stylized 3D scene.
The mean and standard deviation of \(\hat{\mathbf{F}}_\text{VGG}^\text{NF}(\mathbf{r})\) are keep tracked with the moving average during training~\citep{liu2023stylerf}. 

\subsection{Multi-reference image 3D Scene PST via Semantic matching and Local AdaIN}
\label{sec:pst_semantic_matching}
\noindent{With the trained scene content field and AdaIN, we can efficiently transfer the style of images to the 3D radiance field. 
}
However, it often fails to produce satisfactory results when it comes to large-scale 3D scenes:
AdaIN only allows a single reference image which often cannot cover all components in the large-scale scene.
To overcome this limitation, we propose a multi-reference based 3D scene PST by semantically matching the radiance field and multiple reference images.
As shown in the \Fref{fig:pipeline_inference}-(b), the process involves two steps.
First, we compose a style dictionary containing local semantic-style code pairs obtained from semantically 
clustered reference images. 
Then, we perform a semantic-aware style transfer 
by leveraging the semantic correspondence between the 3D scene and each element of the composed style dictionary.

\paragraph{Reference image clustering}
To stylize a 3D scene with multiple style reference images,
we consider a set of reference images, $\mathcal{I}_{s}{=}\{\mathbf{I}_{s}^{i}\}_{i=1,\dots,N}$, where $N$ denotes the number of reference images we use.
We compose a compact style dictionary $\mathcal{D}$ with the reference images by clustering them with similar styles and semantics.
We first extract semantic feature maps to cluster the reference images according to semantic similarity.
We employ DINO~\citep{caron2021emerging} as a semantic feature encoder, which can be generalized to various domains by being trained on large-scale datasets in a self-supervised manner.
We then apply K-means clustering to the extracted semantic feature map \(\mathbf{F}_\text{DINO}(\mathbf{I}_s^i)\), and obtain semantically correlated $M$ number of clusters \(\mathcal{S} {=}\{\mathbf{S}^{ij}\}_{i=1,\dots,N}^{j=1,\dots,M} \) from each reference image $\mathbb{I}_{s}^{i}$.

We obtain local style codes from the clusters by extracting another feature map \(\mathbf{F}_\text{VGG}(\mathbf{I}_s^i) \) from each reference image with VGGNet~\citep{simonyan2014very}.
Then we obtain the local style code, mean \(\bmu_s^{ij}\) and standard deviation \(\bsigma_s^{ij}\), from VGG features \(\mathbf{F}_\text{VGG}(\mathbf{I}_s^{ij}) \in \mathbf{S}^{ij}\) assigned to each cluster.
The centroid \(\bar{\mathbf{f}}_\text{DINO}(\mathbf{I}_s^{ij})\) of each clustered semantic feature and the assigned local style code (\(\bmu_s^{ij}, \bsigma_s^{ij}\))  compose a key-value pair for the style dictionary \(\mathcal{D}\), as \(\mathcal{D} {=} \{ \bar{\mathbf{f}}_\text{DINO}(\mathbf{I}_s^{ij}):( \bmu_s^{ij}, \bsigma_s^{ij})\}_{i=1,\dots,N}^{j=1,\dots,M} \).
With this compact style dictionary, we can efficiently perform local style transfer using multiple reference images by semantically matching the clusters with the 3D scene.

\begin{figure}[t]
    \centering
    \includegraphics[width=\linewidth]{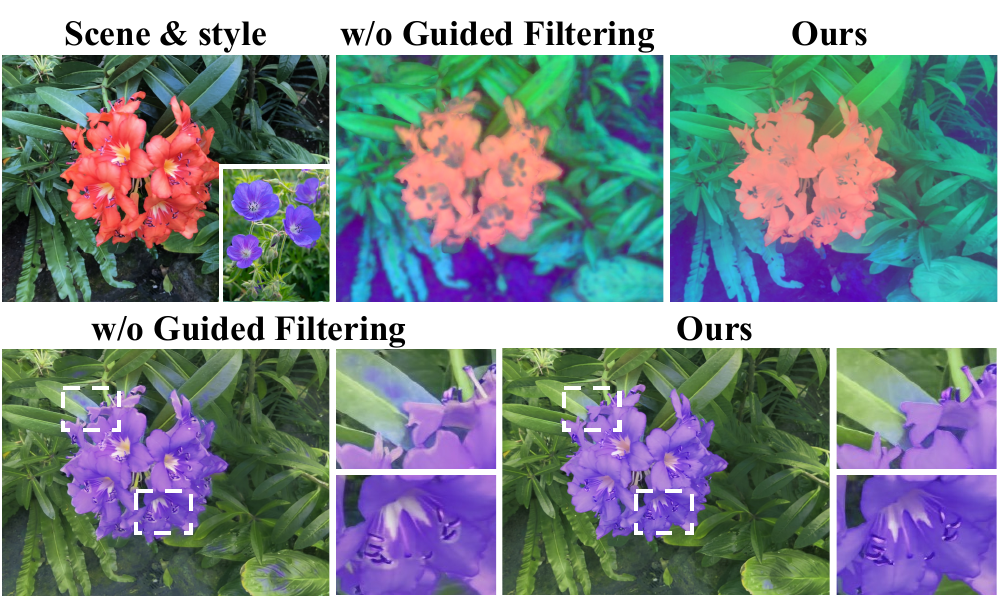} 
    \caption{\textbf{Effects of guided filtering on semantic features.} 
    \textbf{[Top]} Given the trained 3D scene and the reference image (left), we visualize the learned stylizable radiance field using PCA without (mid) / with (right) guided filtering. 
    The learned semantic features are much sharper when guided filtering is applied.
    \textbf{[Bottom]} The stylizable radiance field    shows degraded stylization results if learned without guided filtering, \eg, blurry boundaries (left), higher stylization quality when learned with guided filtering (right).}
    \label{fig:guided_filtering_ablation}
\end{figure}

\paragraph{Scene semantic field}
To semantically match the 3D scene and the reference image clusters, we design and learn an auxiliary 
3D feature grid, called scene semantic field.
The scene semantic field contains semantic features of the 3D scene~\citep{kobayashi2022decomposing, tschernezki2022neural, kerr2023lerf}.
To distill semantic features to the scene semantic field, we optimize the tri-plane features and MLP (orange grid and MLP in \Fref{fig:pipeline_train}) by minimizing the error between rendered features \(\hat{F}_\text{DINO}(\mathbf{r})\) and features extracted from the input images by DINO as follows:
\begin{equation}
    \mathcal{L}_\text{semantic} = \sum_{\mathbf{r} \in \mathcal{R}} \lVert \hat{\mathbf{F}}_\text{DINO}^\text{NF}(\mathbf{r}) - \mathbf{F}_\text{DINO}(\mathbf{I}, \mathbf{r})\rVert_1,
    \label{eq:loss_semantic}
\end{equation}
where \(\mathbf{F}_\text{DINO}(\mathbf{I}, \mathbf{r})\) denotes ground truth DINO features matched to ray \(\mathbf{r}\).
The density $\sigma$ from the scene content field is used for volume rendering, and \(\mathcal{L}_\text{semantic}\) does not affect the learning of the density.
We do not query view direction as input, in order to preserve multi-view semantic consistency. 
Also, for constructing a fine-grained scene semantic field, the ground truth DINO feature maps are refined by guided filtering~\citep{he2012guided}.
The guided filtering enables consistent distillation of semantic features, resulting in clean and photorealistic stylized outputs (see \Fref{fig:guided_filtering_ablation}).
%


\paragraph{Semantic correspondence matching \& Local AdaIN}
When rendering the stylized 3D scene, we use two semantic feature matrices for computing semantic correspondence between the 3D scene and the elements of the style dictionary $\mathcal{D}$. 
The first one is 
\(\hat{\mathbf{F}}_\text{DINO}^\text{NF} \in \mathbb{R} ^{K\times C_{D}}\), obtained from the scene semantic field, where \(K\) is the number of queried 3D points, and \(C_{D}\) is the channel size of the semantic feature, \ie, DINO feature.
The other one is \(\bar{\mathbf{F}}_\text{DINO}(\mathbf{I}_s) \in \mathbb{R} ^{T \times C_{D}} \) comprising keys \(\bar{\mathbf{F}}_\text{DINO}(\mathbf{I}_s^{ij})\) of \(\mathcal{D}\), which are the centroids of the \(T\) clusters.
Given semantic feature matrices, \(\hat{\mathbf{F}}_\text{DINO}^\text{NF}\) and \(\bar{\mathbf{F}}_\text{DINO}(\mathbf{I}_s)\), we compute a cross-correlation matrix \(\mathbf{R}\) as:
\begin{equation}
    \mathbf{R} = \hat{\mathbf{F}}_\text{DINO}^\text{NF}\,\bar{\mathbf{F}}_\text{DINO}(\mathbf{I}_s)^{\top}.
    \label{eq:cross correlation}
\end{equation}

To map local styles of the reference images according to \(\textbf{R}\), we compose two matrices, \(\textbf{M}(\mathbf{F}_\text{VGG}(\mathbf{I}_s)) \in \mathbb{R} ^{T \times C_{V}}\) and \( \mathbf{\Sigma}(\mathbf{F}_\text{VGG}(\mathbf{I}_s)) \in \mathbb{R} ^{T \times C_{V}}\),
comprising the local style codes (\(\bmu_s^{ij}\), \(\bsigma_s^{ij}\)) from the style dictionary,
where \(C_{V}\) denotes the channel size of VGG feature map.
We compute the matrix form of semantic-weighted style codes $(\textbf{M}_{w}, \mathbf{\Sigma}_w)$ as follows:
\begin{equation}
    \begin{gathered}
    \textbf{M}_w = \mathbf{R}^\texttt{S} \textbf{M}(\mathbf{F}_\text{VGG}(\mathbf{I}_s)), \  \mathbf{\Sigma}_w = \mathbf{R}^\texttt{S} \mathbf{\Sigma}(\mathbf{F}_\text{VGG}(\mathbf{I}_s)),
    \label{eq:mw_sw}
    \end{gathered}
\end{equation}
where \(\mathbf{R}^\texttt{S} = \texttt{Softmax}(\mathbf{R})\) demonstrates style attention assigned to the queried 3D point features.
The semantic-weighted style code (\(\textbf{M}_w\), \(\mathbf{\Sigma}_w\)) is assigned to each 3D point according to the style attention.
We feed these semantic-weighted style codes to AdaIN layer as $\mathbf{\Sigma}_w\frac{(\hat{\mathbf{F}}_\text{VGG}^\text{NF} - \bmu_c)}{\bsigma_c} + \textbf{M}_w
$,
followed by the decoder $D_\text{VGG}$ decoding the stylized features into colors.
Finally, we perform volumetric rendering to obtain the final stylized scene.
Note that this semantic-weighted local AdaIN preserves the multi-view stylized color consistency by directly measuring semantic correspondence between reference images and features on the scene semantic fields~\citep{kobayashi2022decomposing}.
 
We found that computing \(\textbf{M}_w\) and \(\mathbf{\Sigma}_w\) on the feature map resolution without clustering~\citep{gunawan2023modernizing}
is inefficient for the volume rendering framework where iterative rendering of rays is inevitable.
Our clustering enables efficient style transfer, especially for cases using multiple reference images.
We highlight that photorealistic scene stylization results are obtained when ${\sim}10$ clusters are used. 
It is worth noting that 
the clustering process takes no more than 1 \emph{sec.} for each reference image. 
Using the small number of clusters allows us to avoid iterative concatenation of high-dimensional reference image features and effectively reduces the computational cost of matrix multiplication.

\section{Feed-Forward Photorealistic Style Transfer of 3D Gaussians - FPGS}\label{sec:FPGS} 

\noindent In this section, we propose \ours, a feed-forward photorealistic style transfer method of scenes represented by 3D Gaussians~\citep{kerbl3Dgaussians}.
FPGS is an extended version of FPRF, which achieves real-time rendering by redesigning the style transfer process of FPRF and leveraging Gaussian Splatting.
In \Sref{sec:feature embedding}, we propose a novel feature field representation by using an RGB to high-dimensional feature mapping network. 
In \Sref{sec:faster style transfer}, we elaborate on algorithmic enhancements of FPGS from FPRF, in \Sref{sec:iterative style transfer}, we propose an iterative style transfer method enhancing local style transfer quality.

\subsection{
Feature fields with 3D Gaussians
}\label{sec:feature embedding}
\noindent
In FPGS, we represent a target scene with explicit 3D Gaussians containing high-dimensional content features and semantic features, which correspond to the scene content field and the scene semantic field of FPRF.
Similar to \Eref{eq:feature volume rendering}, high-dimensional features on 3D Gaussians can be rendered by modifying \Eref{eq:3dgs color volume rendering} as:
\begin{equation}
\hat{\mathbf{F}}(\mathbf{r}) = \sum_{\textbf{g}_i \in \mathcal{G}_\mathbf{r}}
{\hat{\mathbf{f}}}_i\alpha^\text{GS}_i\prod^{i-1}_{j=1}(1-\alpha^{\text{GS}}_j),
\label{eq:3dgs feature volume rendering}
\end{equation}
where ${\hat{\mathbf{f}}}_i$ is the high-dimensional feature assigned to each 3D Gaussian $\textbf{g}_i$.
We can train feature fields represented with 3D Gaussians by minimizing error between rendered $\hat{\mathbf{F}}(\mathbf{r})$ and 2D image features extracted from input images.
However, directly assigning high-dimensional features to 3D Gaussians is  memory intensive and rendering high-dimensional features is time-consuming~\citep{qin2024langsplat, liu2024stylegaussian, zhou2024feature, jun2025dr}.

\paragraph{Scene content field}
To address these inefficiencies, we circumvent the feature
distillation process of the scene content field. 
Different from FPRF which reconstructs the scene content field by distilling features extracted from VGGNet, FPGS first trains the scene content field with RGB loss $\mathcal{L}_\text{RGB}$ only, as shown in the \Fref{fig:pipeline_train}.
Then we employ a mapping network which transforms the RGB values of the 3D Gaussians into high-dimensional features which follow VGG feature distribution.
By leveraging the mapping network as an upsampling network, we can implement the scene content field without feature distillation, which requires memory-intensive assignments of high-dimensional features on 3D Gaussians.
However, we cannot utilize the original VGGNet as a mapping network, whose 2D CNN architecture is not compatible with mapping the RGB values of Gaussians in 3D space.

\begin{figure}[t]
    \centering
    \includegraphics[width=\linewidth]{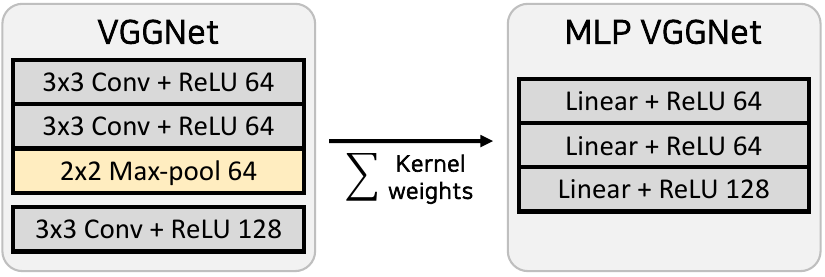}
    \caption{\textbf{Architecture of the MLP VGGNet.} 
    We convert the original VGGNet to MLP VGGNet by mimicking computation of CNN in homogeneous region. The output of the MLP VGGNet approximates the original VGG features and enables style transfer with AdaIN by mapping RGB values to VGG features.
    }
    \label{fig:cnn_to_mlp}
\end{figure}

To alleviate this issue, we propose an MLP VGGNet, which maps a single RGB value of each Gaussian to a VGG feature by interpreting the computation of VGGNet's CNN architecture with an MLP. 
To map a single RGB value to a VGG feature, we can consider a case we predict a VGG feature map from an image containing only a single RGB value. 
In this case, the same input value is multiplied with each weight of convolution kernels and summed. 
As a result, the CNN architecture of VGGNet activates as an MLP, which comprises the same number of neurons as the number of convolution kernels in VGGNet.
The $j$th weight in $i$th layer of the MLP is represented as: 
\begin{equation}
w^{\text{MLP}}_{ij} = \sum_{k=1}^{S_{ij}}{w^{\text{VGG}}_{ijk}},
\label{eq:CNN to MLP}
\end{equation}
where $S_{ij}$ is the kernel size and $w^{\text{VGG}}_{ijk}$ is the kernel weight of $j$th convolution kernel in $i$th layer of VGGNet.
Motivated by this intuitive case example, we convert VGGNet to a novel MLP network called MLP VGGNet.
As shown in \Fref{fig:cnn_to_mlp}, we sum all weights of each kernel in the VGGNet up to the \texttt{ReLU2\textunderscore1} layer, and construct an MLP with the summed weights.
Pooling layers of VGGNet are removed while ReLU layers are maintained after the weight-distillation process.
Using the obtained MLP VGGNet, we represent the scene content field by predicting VGG features with the all RGB values obtained from the pre-optimized 3D Gaussians.
The content features from the MLP VGGNet are compatible with the decoder $D_\text{VGG}$ trained with the original VGGNet, which enables style transfer with 3D Gaussians using AdaIN. 
The MLP VGGNet enables the construction of the scene content field without high-dimensional feature distillation process during scene optimization.

This weight-distilled MLP VGGNet effectively predicts the original VGG feature from a single RGB value.
The original VGGNet activates identically to the MLP VGGNet when the input image contains only a single RGB value, which means the receptive field of the target pixel is a homogeneous area.
In our method, we approximate features from the shallow \texttt{ReLU2\textunderscore1} layer, which has relatively small receptive fields.
The smaller receptive fields hold higher homogeneity, which allows the MLP VGGNet to accurately approximate the original VGG features from the shallow layers.
Note that the proposed converting method does not require any additional training or a fine-tuning process to obtain the MLP VGGNet, so that we can leverage pre-trained VGGNet instantly. 

\paragraph{Scene semantic field}
We cannot employ the same weight-distilled MLP for representing the scene semantic field since semantic encoders require large receptive fields to acquire semantic information from the entire image.
Inspired by Langsplat~\citep{qin2024langsplat}, we train a scene-specific semantic feature autoencoder for compression, which maps high-dimensional features to low-dimensional latent space, and upsamples again to the high-dimensional features.
Before scene reconstruction,
we extract semantic feature maps $\mathbf{F}_\text{DINO}(\mathbf{I})$ from the training images 
$\mathcal{I}$ 
with DINO, and train the autoencoder with:
\begin{multline}
\mathcal{L_\text{AE}} = \sum_{\mathbf{I} \in \mathcal{I}}|| D_\text{AE}(E_\text{AE}(\mathbf{F}_\text{DINO}(\mathbf{I}))) - \mathbf{F}_\text{DINO}(\mathbf{I})||^2_2 \\ 
+ \lambda_{\text{cos}}\sum_{\mathbf{I} \in \mathcal{I}}(1 - \mathtt{sim}(D_\text{AE}(E_\text{AE}(\mathbf{F}_\text{DINO}(\mathbf{I}))), \mathbf{F}_\text{DINO}(\mathbf{I}))),
\label{eq:Autoencoder loss}
\end{multline}
where $E_\text{AE}$ and $D_\text{AE}$ denote the encoder and decoder of the autoencoder, and
$\mathtt{sim}(\cdot, \cdot)$ represents the similarity measure, which in this case is cosine similarity.
With the trained encoder, we encode high-dimensional feature maps $\mathbf{F}_\text{DINO}(\mathbf{I})$ to low-dimensional feature maps, and distill them to 3D Gaussians with \Eref{eq:3dgs feature volume rendering} during the scene optimization.
After optimization, the stylizable 3D Gaussians can be parameterized as: $\mathcal{G}=\{\textbf{g}_i = (\mathbf{p}_i, \mathbf{\Sigma}_i, \sigma_i, \mathbf{K}_i, \hat{\mathbf{f}}_i^\text{GS})\}_{i=1,...,P}$, where $\hat{\mathbf{f}}_i^\text{GS}$ is a distilled low-dimensional semantic feature.
In stylization process, we decode $\hat{\mathbf{f}}_i^\text{GS}$ to a high-dimensional semantic feature $(\hat{\mathbf{f}}_\text{DINO}^\text{GS})_i$ and use it
for semantic matching.
The decoder $D_\text{AE}$ and the MLP VGGNet activate as upsampling networks which map low-dimensional semantic features and RGB values to high-dimensional features, enabling reconstruction of high-dimensional feature fields with 3D Gaussians.

\subsection{
Holistic scene style transfer and real-time rendering
}\label{sec:faster style transfer}
\noindent
In contrast to FPRF, \ours~utilizes an explicit representation for scene reconstruction and editing.
FPRF uses an implicit representation to reconstruct feature fields, which requires query point sampling and a point-wise inference process to obtain high-dimensional features from 3D space.
To stylize the scene using the queried high-dimensional features, the stylization process is needed for every new view rendering, which requires extensive computational cost and slows down the rendering speed.

To overcome this limitation, as shown in \Fref{fig:pipeline_inference}, we decouple the stylization process and the rendering process of FPGS by leveraging explicit nature of 3D Gaussians, which enables real-time rendering. 
We first decode the low-dimensional semantic features contained in all $P$ 3D Gaussians to the high-dimensional semantic features $\hat{\mathbf{F}}_\text{DINO}^\text{GS} \in \mathbb{R} ^{P\times C_{D}}$, using $D_\text{AE}$.
The decoded explicit 3D features from Gaussians enable semantic matching between the entire scene and reference images $\textbf{I}_s$ using modified \Eref{eq:cross correlation} and \Eref{eq:mw_sw} 
as: 
\begin{equation}
    \begin{gathered}
    \mathbf{R} = \hat{\mathbf{F}}_\text{DINO}^\text{GS}\,\bar{\mathbf{F}}_\text{DINO}(\mathbf{I}_s)^{\top} \\
    \textbf{M}_w = \mathbf{R}^\texttt{S} \textbf{M}(\mathbf{F}_\text{VGG}(\mathbf{I}_s)), ~~ \mathbf{\Sigma}_w = \mathbf{R}^\texttt{S} \mathbf{\Sigma}(\mathbf{F}_\text{VGG}(\mathbf{I}_s)).
    \label{eq:FPGS semantic matching}
    \end{gathered}
\end{equation}
The semantic-weighted style code (\(\textbf{M}_w\), \(\mathbf{\Sigma}_w\)) is assigned to all 3D Gaussians, which contains semantically corresponding styles from the reference images $\mathbf{I}_s$.
To stylize the whole 3D Gaussians with the obtained style code, we acquire the scene content features $\hat{\mathbf{F}}^\text{GS}_\text{VGG} \in \mathbb{R} ^{P\times C_{V}}$, by encoding base colors ${\mathbf{C}}^\text{diff} = [\mathbf{c}_1^{\text{diff}}, \mathbf{c}_2^{\text{diff}}, ... ,\mathbf{c}_P^{\text{diff}}]^\top \in \mathbb{R} ^{P\times 3}$ of 3D Gaussians (see \Sref{sec:3dgs preliminary}) with the MLP VGGNet.
Then we stylize the 3D Gaussians using AdaIN as:
\begin{equation}
    \begin{gathered}
    (\hat{\mathbf{F}}^\text{GS}_\text{VGG})' = \mathbf{\Sigma}_w\frac{(\hat{\mathbf{F}}^\text{GS}_\text{VGG} - \bmu_c)}{\bsigma_c} + \textbf{M}_w\\
    (\mathbf{C}^\text{diff})^{'} = D_\text{VGG}((\hat{\mathbf{F}}^\text{GS}_\text{VGG})'),
    \label{eq:gaussian style transfer}
    \end{gathered}
\end{equation}
where $\bmu_c$ and $\bsigma_c$ are the mean and standard deviation of $\hat{\mathbf{F}}^\text{GS}_\text{VGG}$, and $(\mathbf{C}^\text{diff})^{'} \in \mathbb{R} ^{P\times 3}$ denotes the stylized base color of 3D Gaussians.
With the stylized base color, we update SH coefficients $\mathbf{K}_i$ of each 3D Gaussian $\textbf{g}_i$ using \Eref{eq:spherical harmonics} as: $(\mathbf{K}_i^{[:,1:2]})' = (\mathbf{c}^\text{diff}_{i})^{'} / C_\text{SH}$.
This process stylizes the entire scene at once in a feed-forward manner. 
After we obtain the stylized Gaussians, we can render the stylized scene in real-time without rendering the high-dimensional features.

\subsection{
Iterative style transfer
}\label{sec:iterative style transfer}
\noindent
To enhance local style transfer quality, we propose an iterative style transfer method.
By decoupling the style transfer and the rendering process, we can perform additional style transfer process without compromising rendering speed. 
After obtaining semantic-weighted style code $(\textbf{M}_w, \mathbf{\Sigma}_w)$ by semantic correspondence matching, we perform local AdaIN iteratively as:
\begin{equation}
    \begin{gathered}
    (\hat{\mathbf{F}}^\text{GS}_\text{VGG})^i = E_\text{VGG}((\mathbf{C}^\text{diff})^{i})\\
    (\hat{\mathbf{F}}^\text{GS}_\text{VGG})^{i+1} = \mathbf{\Sigma}_w\frac{((\hat{\mathbf{F}}^\text{GS}_\text{VGG})^i - \bmu^i_c)}{\bsigma^i_c} + \textbf{M}_w\\
    (\mathbf{C}^\text{diff})^{i+1} = D_\text{VGG}((\hat{\mathbf{F}}^\text{GS}_\text{VGG})^{i+1}),
    \label{eq:iterative style transfer}
    \end{gathered}
\end{equation}
where $(\hat{\mathbf{C}}^\text{diff})^{i}$ and $(\hat{\mathbf{F}}^\text{GS}_\text{VGG})^{i}$ denotes the diffuse colors and content features at $i$th iteration, $\bmu_i$ and $\bsigma_i$ are the mean and standard deviation of $(\hat{\mathbf{F}}^\text{GS}_\text{VGG})^{i}$. 
In each style transfer process, the content features are normalized by the global style code $(\bmu^i_c, \bsigma^i_c)$ and transformed with the semantic-weighted local style code $(\textbf{M}_w, \mathbf{\Sigma}_w)$. 
By updating the style of the scene iteratively, we can enhance the locality of the stylization, which transfers the local style to the semantically corresponding regions of the 3D scene more precisely (see \Fref{fig:ablation_studies}).

\begin{figure*}[t]
    \centering
        \includegraphics[width=\linewidth]{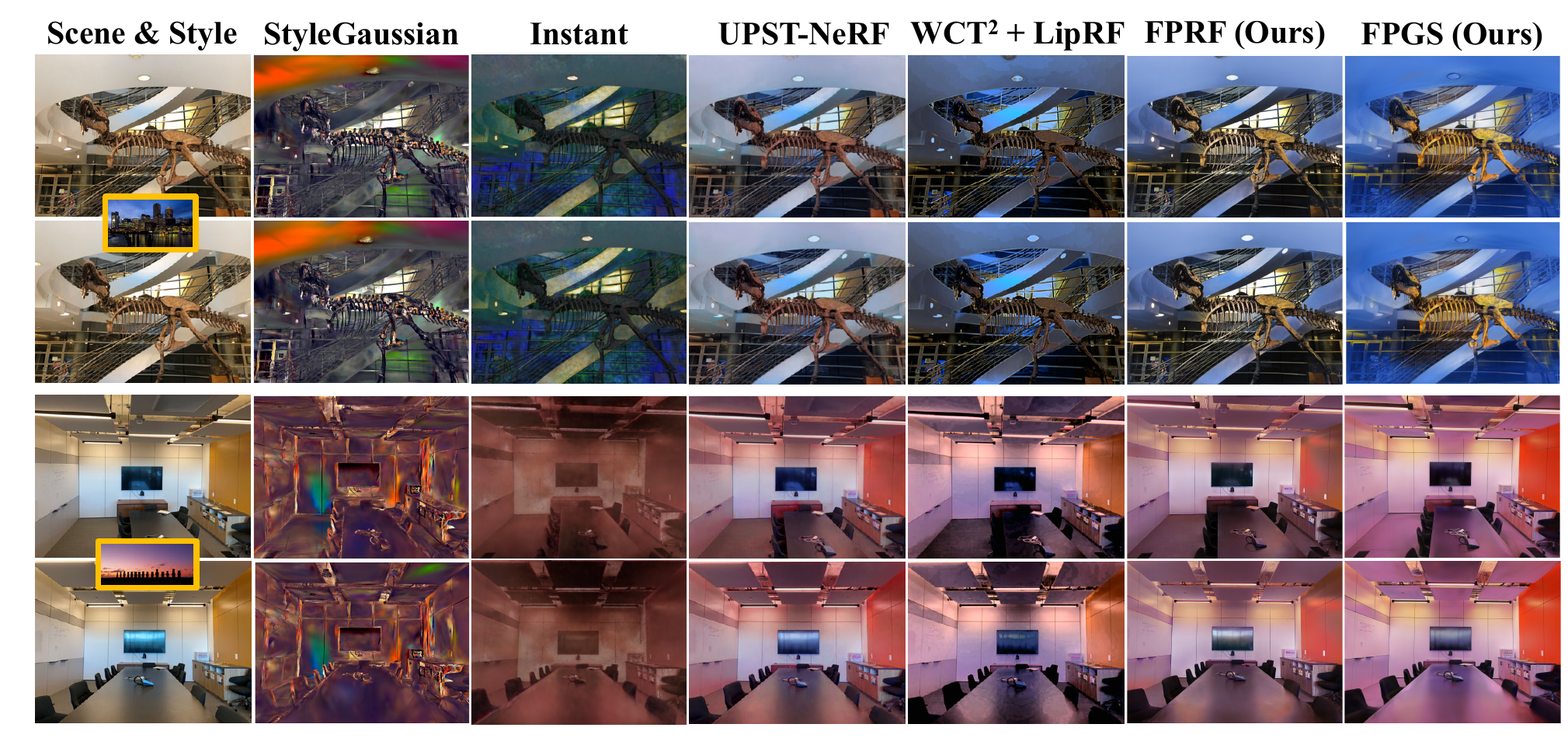}
        \caption{\textbf{PST quality comparison on the LLFF dataset~\citep{mildenhall2019local}.}
        Compared to the competing 3D PST methods, our methods stylize the radiance field in a photorealistic manner by transferring the diverse color of the reference image while preserving the original images’ naturalness and vividness.
        \textbf{Instant} denotes Instant-NeRF-Stylization~\citep{li2024instant} method, and $\textbf{WCT}^2 + \textbf{LipRF}$ denotes LipRF~\citep{zhang2023transforming} based on the 2D PST method, $\textbf{WCT}^2$~\citep{yoo2019photorealistic}
        }
\label{fig:LLFF_comparison}
\end{figure*}

\begin{figure*}[thbp]
    \centering
        \includegraphics[width=\linewidth]{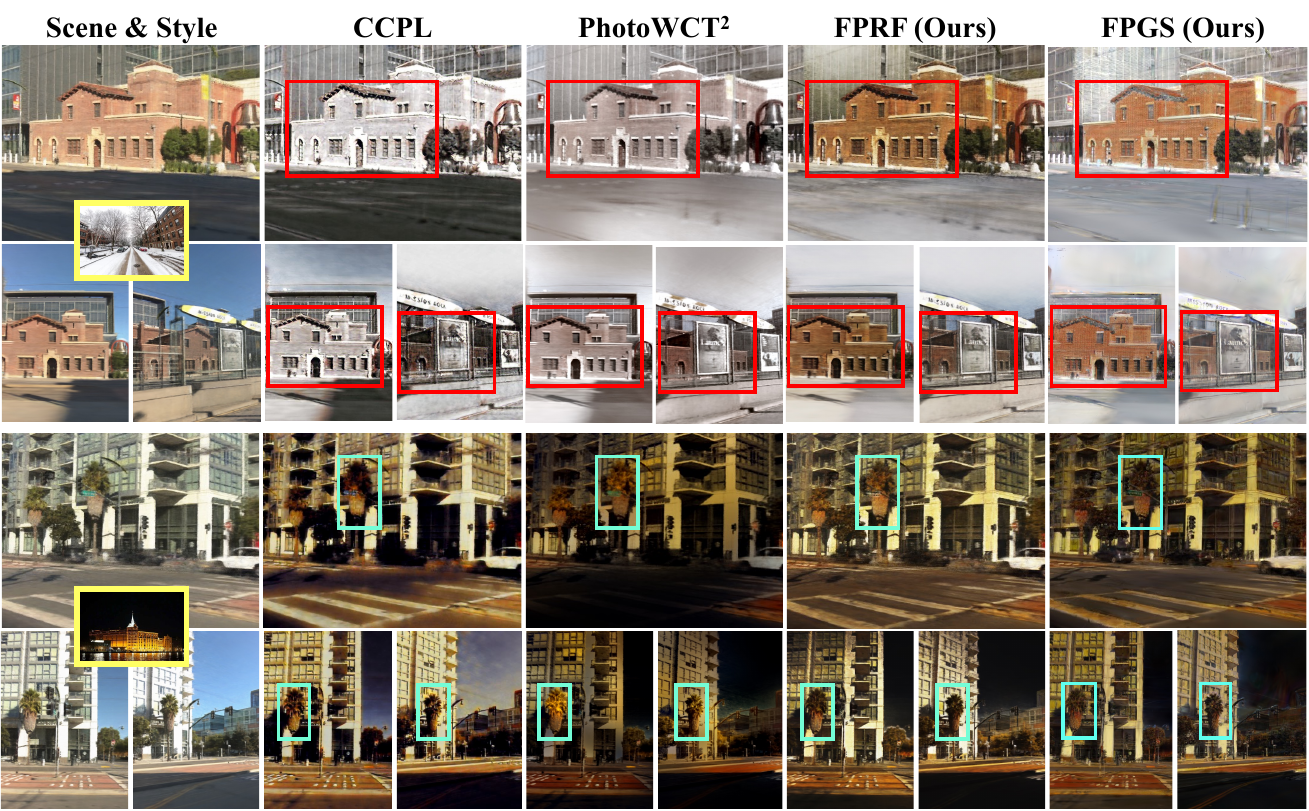}
    \caption{
   \textbf{Multi-view appearance consistency on the San Francisco Mission Bay dataset~\citep{tancik2022blocknerf}.}
   Our methods preserve multi-view appearance consistency even in extreme viewpoint change, while 2D PST methods~\citep{wu2022ccpl, chiu2022photowct2} produce inconsistent colors of the same building as the viewpoint changes. }
\label{fig:large_scale_comparison}
\end{figure*}

\section{Experiments}
\label{sec:experiment}
\noindent
\paragraph{Datasets}
We evaluate our methods on two public datasets, the LLFF~\citep{mildenhall2019local} dataset for small scenes and the San Fran Cisco Mission Bay dataset~\citep{tancik2022blocknerf} for large-scale scenes.  
The LLFF dataset includes 8 real forward facing scenes and the San Fran Cisco Mission Bay dataset is a city scene dataset consisting of about 12,000 images recorded by 12 cameras.
For 4D scene experiments, we use KITTI-360~\citep{liao2022kitti}  comprising 9 urban driving sequences, and we also use Tanks and Temples~\citep{knapitsch2017tanks} containing unbounded scenes for additional experiments.

\subsection{Qualitative results}
\paragraph{PST on small-scale scenes}
We compare our model with three competing 3D PST methods; Instant-NeRF-Stylization~\citep{li2024instant}, 
UPST-NeRF~\citep{chen2024upst}, LipRF~\citep{zhang2023transforming}, and recent feed-forward 3D \emph{artistic} style transfer methods, StyleRF~\citep{liu2023stylerf} and StyleGaussian~\citep{liu2024stylegaussian}.
As shown in Fig.~\ref{fig:LLFF_comparison}, our methods transfer the diverse colors of the style reference image well while maintaining the texture fidelity of the original scene.
While UPST-NeRF effectively retains the structure of the scene, the stylized scene's color style differs from the reference.
LipRF stabilizes the artifacts caused by 2D PST methods~\citep{yoo2019photorealistic, wu2022ccpl} with the Lipschitz network, but it tends to oversmooth the scene and loses the color diversity of the reference image.
Instant-NeRF-Stylization and StyleRF fail to obtain photorealistic results. 
Note that UPST-NeRF, StyleRF, and StyleGaussian require time-consuming per-scene optimization after scene reconstruction for style transfer, LipRF and Instant-NeRF-Stylization need per-style optimization. 
On the contrary, our methods achieve feed-forward PST after an efficient training process.
To compare our methods with LipRF which is not open-sourced, we brought the qualitative results from LipRF paper. In other words, the scenes in \Fref{fig:LLFF_comparison} are not cherry-picked. 

\paragraph{PST on large-scale scenes}
We compare our methods against two competing 2D PST methods, CCPL~\citep{wu2022ccpl} and \(\text{PhotoWCT}^2\)~\citep{chiu2022photowct2}, since we propose the first method aiming for large-scale 3D radiance field PST, no 3D PST method supports large-scale datasets.
Figure~\ref{fig:large_scale_comparison} shows a qualitative comparison on large-scale scenes.
We observe that 2D PST methods fail to preserve multi-view color consistency under wide-range view changes, \eg, they obtain different colors of the same building and sky as the viewpoints change.  
Also, they stylize images without understanding of semantic relation between the content image and the reference image.
Instead, our methods elaborately transfer styles by reflecting the semantic correspondence between scene and reference images.
Semantic matching can also preserve multi-view consistency by directly measuring semantic correspondence between the reference images and the 3D scene semantic field directly. 

More qualitative results from diverse scenes are given in our supplementary material, which shows the generalizability of our methods.

\definecolor{tabfirst}{rgb}{1, 0.7, 0.7} 
\definecolor{tabsecond}{rgb}{1, 0.85, 0.7} 
\definecolor{tabthird}{rgb}{1, 1, 0.7} 

\begin{table}[t]
\centering
\caption{Multi-view consistency comparison. [Top] Comparison with feed-forward 3D style transfer methods. [Bottom] Comparison with 2D PST methods.
Short denotes short baseline and long denotes wide baseline of camera settings. 
}
\resizebox{\linewidth}{!}{
\begin{tabular}{c|cccc}
\toprule
\multicolumn{5}{c}{LLFF dataset}\\
\midrule
 &StyleGaussian & UPST-NeRF  &  FPRF (Ours) & FPGS (Ours)  \\
\midrule
 Short ($\downarrow$) & 0.0264 &  0.0234 & \cellcolor{tabsecond}0.0231 & \cellcolor{tabfirst}0.0214 \\  
Long ($\downarrow$) & 0.0440 & \cellcolor{tabsecond}0.0384 & 0.0399 & \cellcolor{tabfirst}0.0349  \\
\midrule
\midrule
\multicolumn{5}{c}{San Fran Cisco Mission Bay dataset}\\
\midrule
&
CCPL & {$\text{PhotoWCT}^2$} & FPRF (Ours) & FPGS (Ours)  
\\
\midrule
Short ($\downarrow$) &
0.0528 & 0.0525 & \cellcolor{tabsecond}0.0395  & \cellcolor{tabfirst}0.0335 
\\  
Long ($\downarrow$) &
0.0775 & 0.0755 & \cellcolor{tabsecond}0.0563  & \cellcolor{tabfirst}0.0502  
\\
\bottomrule
\end{tabular}
\label{tab:warp_error_block}
}
\end{table}

\definecolor{tabfirst}{rgb}{1, 0.7, 0.7} 
\definecolor{tabsecond}{rgb}{1, 0.85, 0.7} 
\definecolor{tabthird}{rgb}{1, 1, 0.7} 

\vspace{-0.5cm}
\begin{table}[t]
\centering
    \caption{
    Running time comparison with feed-forward 3D style transfer methods on the LLFF dataset~\cite{mildenhall2019local}.
    FPGS requires less training time than 
    the other methods
    and achieves real-time rendering.
    }
    \resizebox{\linewidth}{!}{



        

    \begin{tabular}{c|c|c|c|c}
        \toprule
        & \multicolumn{3}{c|}{Train (min) } & \multirow{2}{*}{Render (sec)} \\
        \cmidrule{2-4}
        & pre-/post-optimization & scene recon. & \textbf{total} & \\
        \midrule
        StyleRF & 301 & 40 & 341 & 16.62 \\
        StyleGaussian & 347 & \cellcolor{tabsecond}11 & 358 & \cellcolor{tabfirst}0.004 \\
        {UPST-NeRF} & 650 & \cellcolor{tabfirst}6 & 656 & \cellcolor{tabsecond}1.357 \\
        {FPRF (Ours)} & \cellcolor{tabfirst}0 & 61 & \cellcolor{tabsecond}61 & 3.812 \\ 
        {FPGS (Ours)} & \cellcolor{tabsecond}4 & 20 & \cellcolor{tabfirst}24 & \cellcolor{tabfirst}0.004 \\
        \bottomrule
    \end{tabular}
    }
    \label{tab:running_time}
\end{table}
\definecolor{tabfirst}{rgb}{1, 0.7, 0.7} 
\definecolor{tabsecond}{rgb}{1, 0.85, 0.7} 
\definecolor{tabthird}{rgb}{1, 1, 0.7} 

\begin{table}[t]
\centering
    \caption{
    User Study on the quality of PST with 3D PST methods. We evaluate on the LLFF~\cite{mildenhall2019local} dataset and
    the rating is of scale 1-5. \textbf{Instant} denotes Instant-NeRF-Stylization~\cite{li2024instant} method.
    }
    \resizebox{\linewidth}{!}{
    \begin{tabular}{c|cccc}
        \toprule
        & Instant & {UPST-NeRF} &{FPRF (Ours)} &  {FPGS (Ours)}\\

        \midrule

        PST quality $(\uparrow)$ & 1.305 & 3.279  &  \cellcolor{tabsecond}3.576 &  \cellcolor{tabfirst}3.936 \\

        \bottomrule
        
    \end{tabular}
    
    }
    \label{tab:user_study}
\end{table}

\subsection{Quantitative results}
\noindent
\paragraph{Multi-view consistency}
We report multi-view consistency of our method by following previous 3D style transfer works~\citep{virmaux2018lipschitz, chen2024upst, liu2023stylerf}.
Multi-view consistency denotes a feature that a shared geometry wherein a scene observed from various angles can consistently explain multiple views of the scene.
We evaluate the multi-view consistency of 3D stylization by computing the multi-view error via image warping. 
Specifically, for the two rendered images \(\textbf{I}_\textbf{d} \) and \(,\textbf{I}_{\textbf{d}'}\) from two different views \(\textbf{d}\) and \(\textbf{d}'\), the multi-view error is computed as:
\begin{equation}
    E_\text{warp}(\textbf{I}_\textbf{d}, \textbf{I}_{\textbf{d}'}) = \text{RMSE}(\textbf{I}_\textbf{d}, W(\textbf{I}_{\textbf{d}'}); \textbf{B}_{\textbf{d}'\textbf{d}}),
\end{equation}
where 
\(W\) warps \(\textbf{I}_{\textbf{d}'}\) to \(\textbf{I}_\textbf{d}\) and $\textbf{B}_{\textbf{d}'\textbf{d}}$ is the binary mask of valid pixels warped from the view \(\textbf{d}'\) to \(\textbf{d}\). The warping function \(W\) and binary mask $\textbf{B}_{\textbf{d}'\textbf{d}}$ are measured from the rendered original scene images, with an off-the-shelf optical flow method~\citep{teed2020raft}. The non-valid pixels are masked out and are not considered. 

We compare our method with feed-forward 3D style transfer methods~\citep{liu2024stylegaussian,chen2024upst} on the 8 scenes of the LLFF dataset~\citep{mildenhall2019local}, and with 2D PST methods~\citep{wu2022ccpl,chiu2022photowct2} on the 4 blocks from the San Fran Cisco Mission Bay dataset~\citep{tancik2022blocknerf}.
We use 20 style images from the PST dataset~\citep{luan2017deep} for evaluation.
We report warp errors in both short-/long-term settings denoting short-/wide-baselines of cameras.
Table \ref{tab:warp_error_block} shows the superior multi-view consistency of FPGS compared to prior methods.
Note that our methods effectively preserve multi-view consistency by computing semantic correspondence with features on the 3D space directly, not on the rendered features.
Our scene semantic field contains view-independent semantic features in the 3D space, which ensures that each 3D point is always stylized with the assigned same local style, regardless of view directions. 

\paragraph{Running time}
\Tref{tab:running_time} shows FPGS significantly outperforms other methods in terms of training and rendering time. 
As training time, we measure the total time the model required to become ready for feed-forward style transfer with arbitrary styles, which includes scene reconstruction time and per-scene pre-/post-optimization time.
Note that scene reconstruction time highly depends on the efficiency of the off-the-shelf  baseline scene representation which can be substituted.
In other words, the key process deciding the training time of feed-forward 3D style transfer model is per-scene post-optimization, which is required to the previous methods~\citep{chen2024upst, liu2023stylerf, liu2024stylegaussian} to obtain generalizability for arbitrary styles.
Our methods effectively circumvent this per-scene post optimization by using pre-trained MLP color decoder enabling style transfer using arbitrary style, without per-scene fine-tuning.
FPGS only requires a short pre-optimization to train a semantic feature autoencoder for compressing the semantic features on 3D Gaussians.

FPGS also achieves real-time rendering by decoupling stylization and rendering process.
In contrast, NeRF-based methods~\citep{chen2024upst, liu2023stylerf, kim2024fprf} intertwine the stylization and rendering processes, which slows down the rendering speed.

\paragraph{User study}
We conduct a user study to evaluate the perceptual quality of stylization.
We evaluate our methods on the LLFF~\citep{mildenhall2019local} dataset by asking 20 volunteers to score (1-5) the stylization quality of 3D PST.
As shown in \Tref{tab:user_study}, FPGS obtained the best score in terms of PST quality. 

\begin{figure}[t]
    \includegraphics[width=\linewidth]{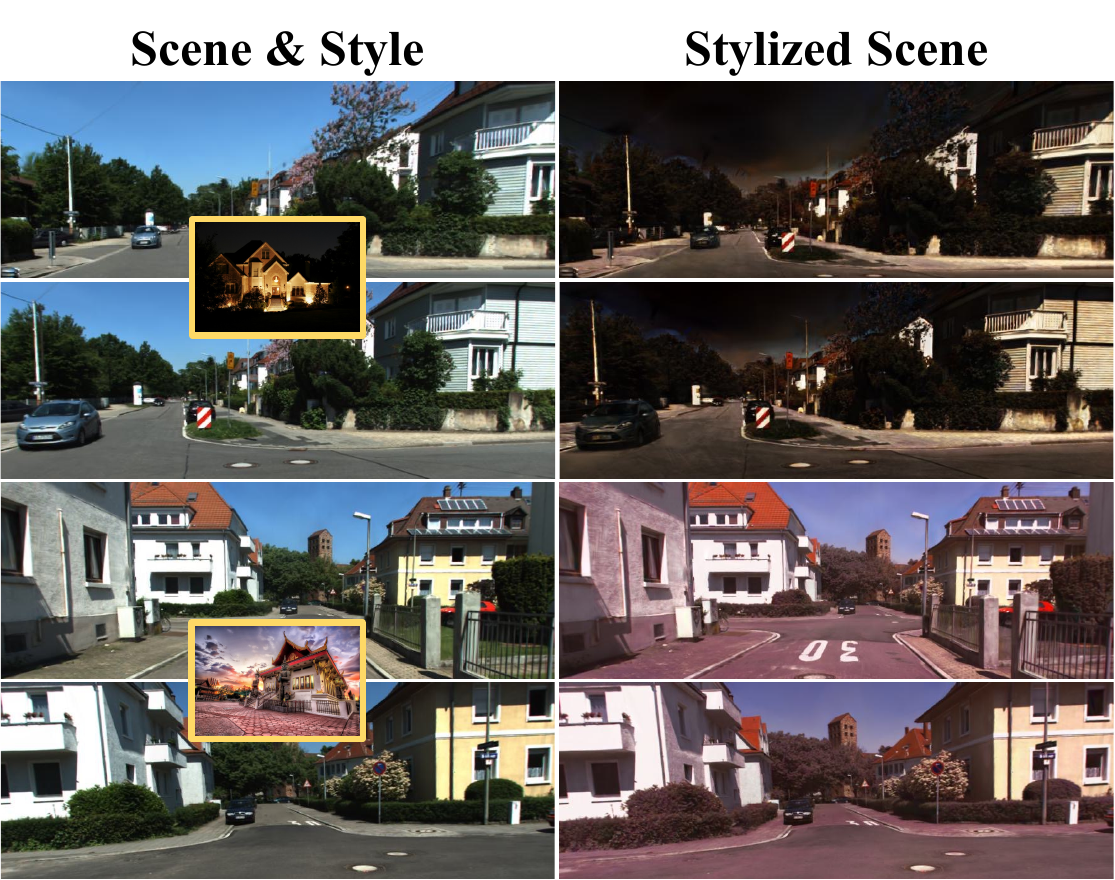} 
    \caption{\textbf{PST on the KITTI-360 dataset~\citep{liao2022kitti} containing 4D urban scenes.} 
    FPGS can be applied to 4D large-scale scenes with time/multi-view consistency.}
    \label{fig:4d_scene}
\end{figure}

\begin{figure}[t]
    \centering
    \includegraphics[width=1\linewidth]{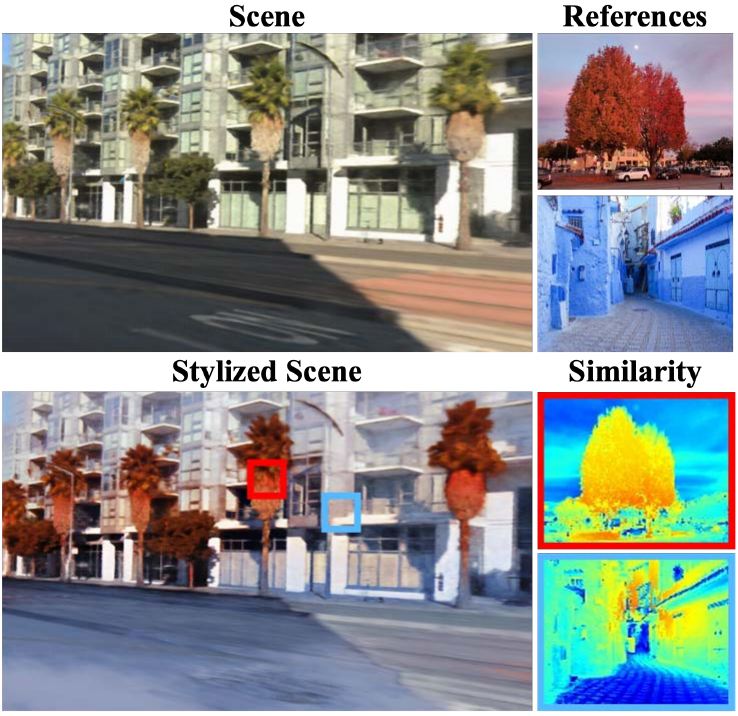} 
    \caption{\textbf{Multi-reference style transfer.}
    \ours~stylizes the 3D radiance field with multiple reference images. Each heatmap shows the similarity between the 
    rendered 
    semantic features from a highlighted patch and the reference image. Our model comprehends the     semantic relationship of a large-scale 3D scene and matches the scene with the reference images.
    }
    \label{fig:multi_reference}
\end{figure}

\begin{figure}[t]
    \includegraphics[width=\linewidth]{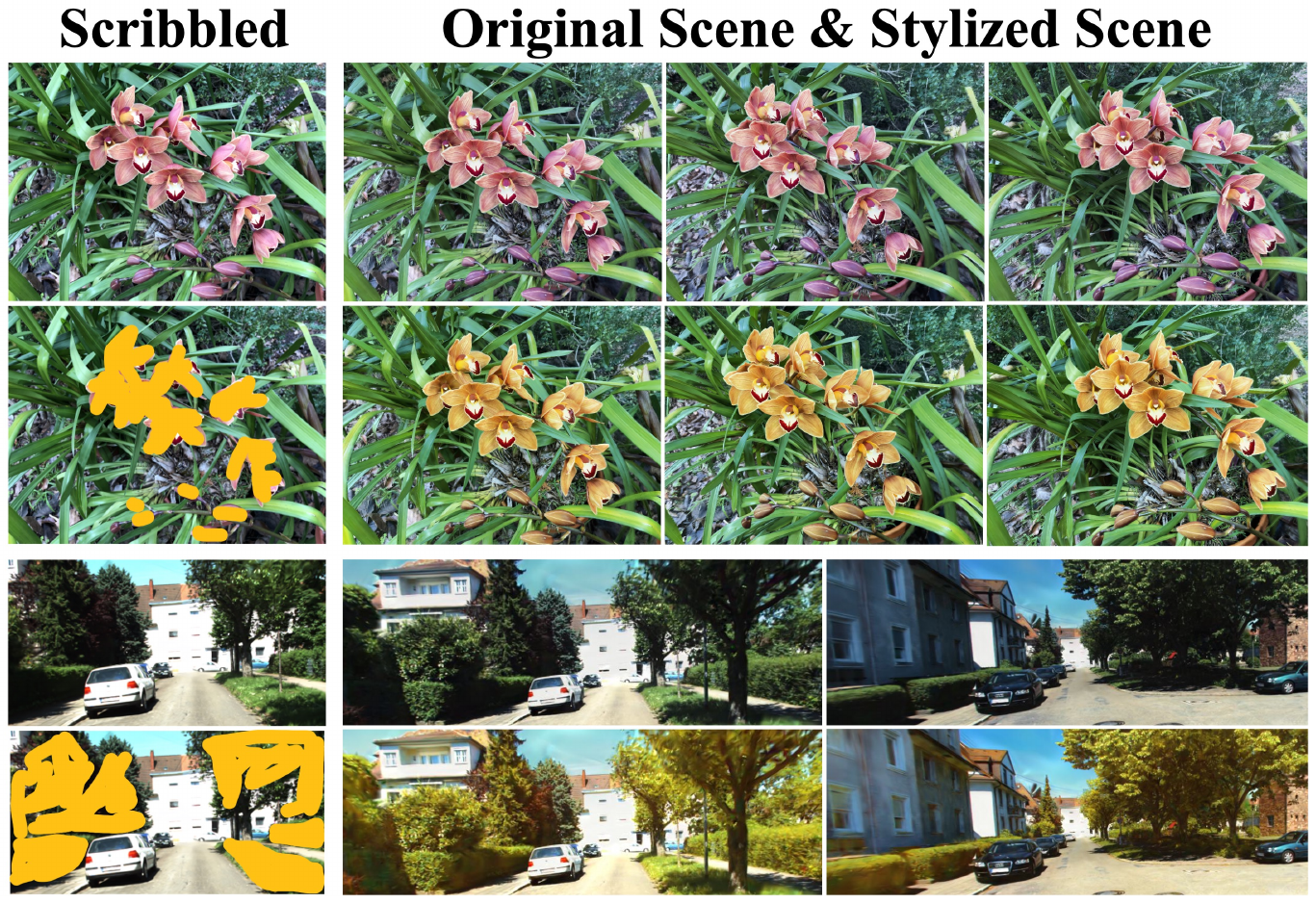} 
    \caption{\textbf{Scribble-based style transfer.} We render images from the optimized scene and draw scribbles on the images. Our methods can transfer the scribbled colors on the images to the corresponding regions of the scene photorealistically.  
    } 
    \label{fig:scribble}
\end{figure}

\subsection{Applications}
\noindent
\paragraph{4D scene style transfer}
We can apply FPGS for stylizing 4D scenes represented by time-variant Gaussians.
To stylize 4D scenes, we reconstruct stylizable 4D scenes based on a 4D Gaussian modeling method, VEGS~\citep{hwang2024vegs}.
VEGS represents a 4D Gaussian $\textbf{g}^t_i$ with a time-variant mean $\mathbf{\bmu}^t_i = \mathbf{T}^t_i\mathbf{\bmu}_i$ and a covariance $ \mathbf{\Sigma}^t_i = \mathbf{R}^t_i\mathbf{\Sigma}_i\mathbf{R}^{t\top}_i$,
where $\mathbf{T}^t_i$ is a transform matrix of $\textbf{g}^t_i$ at timestep $t$, and $\mathbf{R}^t_i$ is the rotation matrix of $\mathbf{T}^t_i$.
We modify the training process of VEGS to jointly optimize colors and semantic features on the 4D Gaussians.
Figure~\ref{fig:4d_scene} shows the rendered images of stylized 4D autonomous driving scenes from the KITTI-360 dataset~\citep{liao2022kitti}, with different timesteps and viewpoints.
As can be seen, the stylized scenes preserve time/multi-view consistency after extreme time/view-point change.
Since we simply embeds semantic features on the time-variant Gaussians, our method inherits time/multi-view consistency from the based 4D reconstruction model.

\paragraph{Multi-reference style transfer}
Figure~\ref{fig:multi_reference} shows the style transfer results using multiple reference images. 
Our methods effectively select suitable styles from multiple references with the scene semantic field. 
Semantic similarity is computed by multiplying features from the scene semantic field with DINO~\citep{caron2021emerging} feature maps extracted from the reference images. 
The similarity map clearly shows that our methods comprehend the semantic relationship between scenes and reference images.

\paragraph{Scribble-based style transfer} 
Figure \ref{fig:scribble} shows that our methods can stylize the scene with scribbles. We slightly modified the semantic matching and local AdaIN process, to match the scribbled colors on the rendered images to semantically corresponding region of the original scene (see details in supplementary material).
The scribbled colors are transferred to the 3D/4D scenes photorealistically, and the results are consistent after extreme changes of viewpoints and timesteps.

\begin{figure}[t]
    \includegraphics[width=\linewidth]{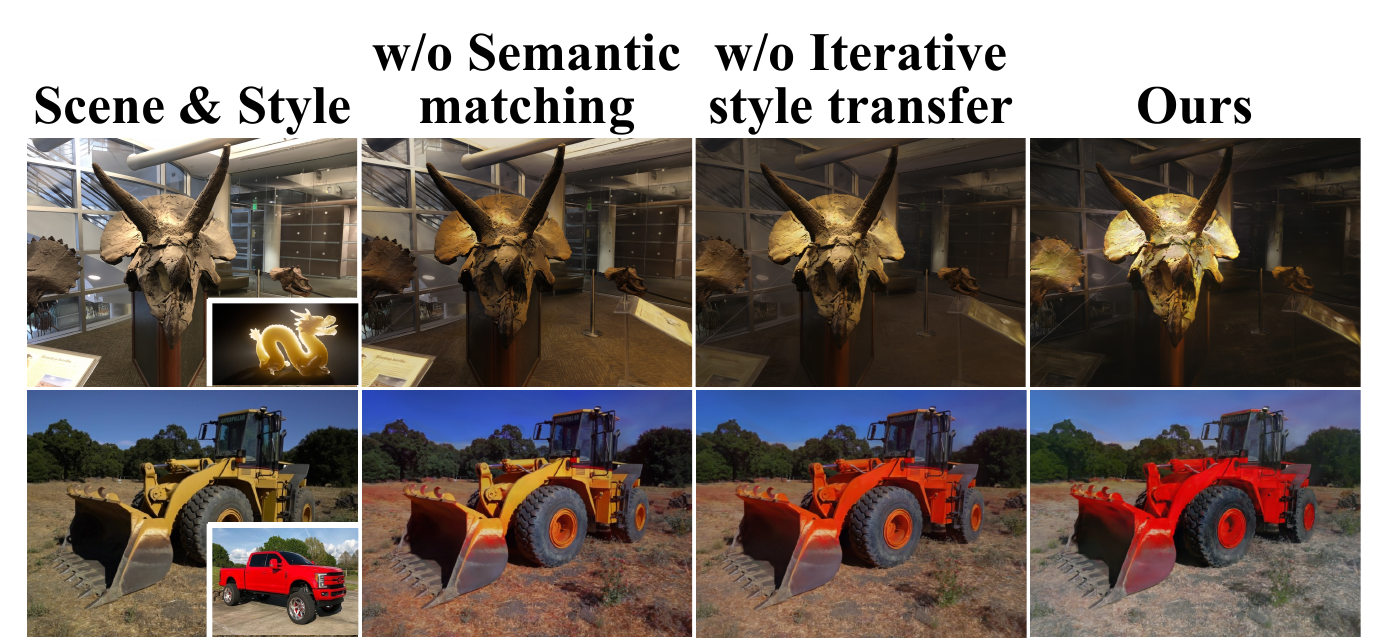} 
    \caption{\textbf{Ablation studies.} 
    “w/o Semantic matching” stylizes the model without semantic matching and local AdaIN, \ie, global style transfer. “w/o Iterative style transfer” performs style transfer with one iteration.}
    \label{fig:ablation_studies}
\end{figure}

\begin{figure}[t]
    \includegraphics[width=\linewidth]{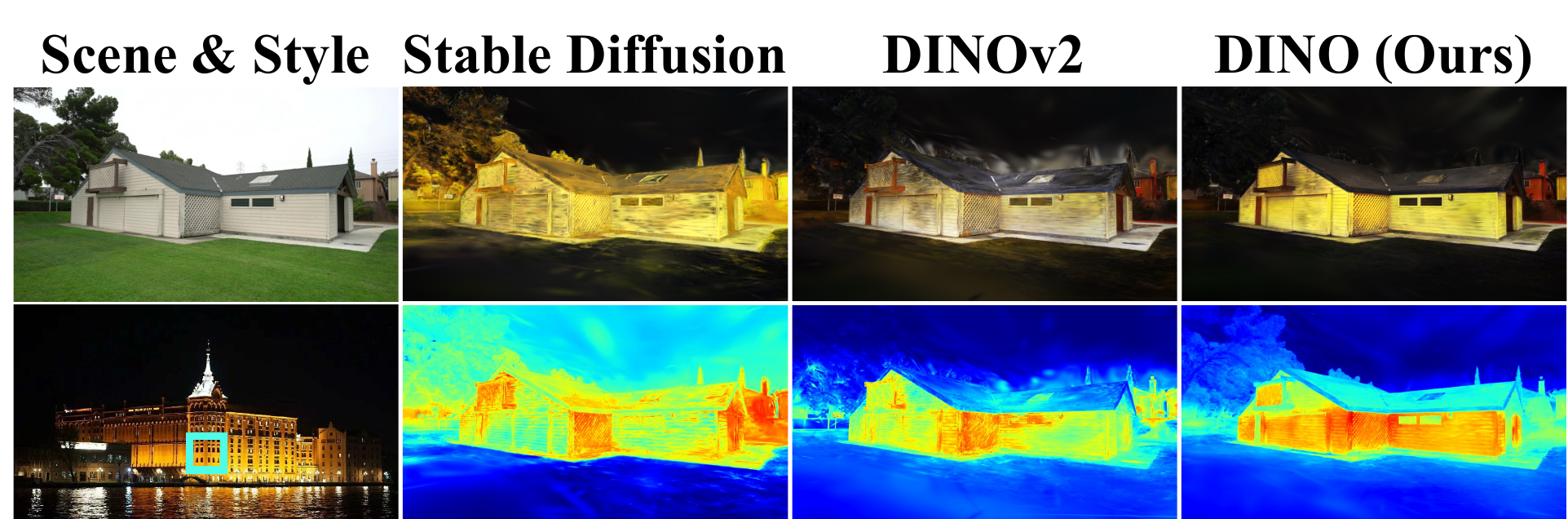} 
    \caption{\textbf{Style transfer results with diverse semantic image encoders.} 
    We construct scene semantic fields  by distilling image features extracted from different image foundation models. Each heatmap shows the similarity between the features extracted from the highlighted patch of the reference image and the semantic features rendered from the scene semantic field.}
    \label{fig:semantic_encoders}
\end{figure}

\subsection{Additional experiments}
\noindent
\paragraph{Ablation studies}
As can be seen in \Fref{fig:ablation_studies}, semantic matching improves perceptual style similarity between the stylized scene and the reference image, by using semantic-weighted local style code for stylization.
Iterative style transfer enhances the effect of local style transfer by iteratively stylizing the scene with the obtained semantic-weighted local style code.

\paragraph{Semantic encoder comparison}
We investigate the effect of semantic encoder selection on the style transfer quality, by training the scene semantic fields with features from the different image foundation models.
We test our method with Stable Diffusion v1-5~\citep{rombach2022high} and DINOv2~\citep{oquab2023dinov2}, which can perform dense semantic correspondence matching with feature distance~\citep{zhang2023tale}.
To utilize Stable Diffusion as a feature encoder, we extract image features from the intermediate layers of denoising U-Net, by following \citep{zhang2023tale}.
As shown in \Fref{fig:semantic_encoders}, the semantic correspondence matching and style transfer quality depends on the semantic features from the foundation
models.
We find DINO~\citep{caron2021emerging} shows the most perceptually satisfactory stylized results and utilize it as our semantic encoder. 

\section{Conclusion}
In this paper, we present FPGS, a novel 3D photorealistic style transfer (PST) method for radiance fields represented by 3D Gaussians. 
FPGS allows PST in a feed-forward manner by leveraging AdaIN, without sacrificing real-time rendering speed inherited from 3D Gaussian representation.
FPGS also supports multi-reference style transfer enabling stylization of large-scale 3D scenes which consist of diverse components.

The current limitation is that the semantic matching performance of our model is bounded by the capability of the semantic image encoder, DINO.
Nonetheless, since our model can utilize any semantic encoder for constructing the scene semantic field, the performance of our model stands to benefit from the emergence of more advanced models.
%


\noindent\textbf{Data availability statements.}
All data supporting the findings of this study are available online.
The LLFF dataset can be downloaded from \url{https://github.com/Fyusion/LLFF}. 
The San Fran Cisco Mission Bay dataset can be downloaded from \url{https://waymo.com/research/block-nerf/}.
The Tank and Temples dataset can be downloaded from \url{https://www.tanksandtemples.org/}.
The KITTI-360 dataset can be downloaded from \url{https://www.cvlibs.net/datasets/kitti-360/}.

{\small
\bibliographystyle{spbasic}
\bibliography{ms}
}

\end{document}


\maketitle


\section*{Appendix A \\ Experimental details}
\subsection{Multi-view consistency comparison}
By following previous works~\citep{huang2022stylizednerf, liu2023stylerf, chen2024upst, hedlin2023unsupervised} we use multi-view error via image warping (see Sec. \uppercase\expandafter{\romannumeral6}-B) as a metric for quantifying multi-view consistency of stylized scenes. 
With this metric, the rendered original scenes
should be same for a fair comparison, since multi-view error of the stylized scene is proportional to multi-view error of the original scene.
However, the original scenes of different methods cannot be same and multi-view errors of the original scenes depend on the capabilities for view-dependent color of comparing methods using different radiance field modelings.

To minimize the multi-view error difference of original scenes rendered from the comparing methods, we disable view-dependent color of each method.
For the NeRF-based~\citep{mildenhall2020nerf} methods, UPST-NeRF~\citep{chen2024upst} and FPRF~\citep{kim2024fprf}, we train and render the scene with a single dummy view direction, which ensures view-independent radiance field reconstruction, \ie, Lambertian model.
For the 3D Gaussian-based~\citep{kerbl3Dgaussians} method, FPGS, we train and render the scene using spherical harmonics of degree 0, which can only represent view-independent radiance field.

We render $K$ images $\{{\textbf{I}_i}\}_{i=1,2,...,K}$ with sampled sequential viewpoints from a continuous camera trajectory and select nearby image pairs with short and long baseline settings to compute warped image errors.
We predict optical flows and binary masks with the rendered original image pairs using RAFT\citep{teed2020raft}, and stylize the images with 20 style images from the PST dataset~\citep{luan2017deep} for evaluation.  

\paragraph{Large-scale scene}
We compare our methods with two competing 2D PST methods~\citep{wu2022ccpl,chiu2022photowct2} on the 4 blocks from the San Fran Cisco Mission Bay dataset~\citep{tancik2022blocknerf}.
We sample 80 image pairs from 4 camera trajectories for each block, which comprise $(\textbf{I}_i, \textbf{I}_{i+1})$ as short baseline pairs and $(\textbf{I}_i, \textbf{I}_{i+3})$ as long baseline pairs.
To compare with 2D methods, we first render the original scenes with FPGS and stylize the rendered images with 2D PST methods.

\paragraph{Small-scale scene}
We evalute our methods with two competing 3D feed-forward methods~\citep{liu2023stylerf,chen2024upst} on the 8 scenes from the LLFF dataset~\citep{mildenhall2019local}.
We sample 24 image pairs from a camera trajectory for each scene, which comprise $(\textbf{I}_i, \textbf{I}_{i+1})$ as short baseline pairs and $(\textbf{I}_i, \textbf{I}_{i+5})$ as long baseline pairs.

\subsection{Running time comparison}
We compare training time and rendering time of our methods with feed-forward 3D style transfer methods, on the LLFF dataset~\citep{mildenhall2019local} (see Tab. \uppercase\expandafter{\romannumeral4}).
To train StyleRF~\citep{liu2023stylerf} and UPST-NeRF~\citep{chen2024upst}, we follow the official implementation.
We train FPRF and FPGS with 30,000 iterations for each scene, and train the semantic autoencoder of FPGS for 5 epochs with the training images.
For a $1008\times756$ resolution scene, we train and render comparing methods on a single NVIDIA RTX A6000 gpu, and obtain the average running time of all 8 scenes.

\subsection{Scribble-based style transfer}
Our methods can stylize the 3D scene with a rendered image and scribbles on the image, as shown in Fig. 9.
The rendered image holds strong semantic correspondence with the projected region from the 3D scene.
This strong semantic correspondence enables transferring of scribbled colors to the matched region in the 3D scene.
We first compute semantic correspondence between the rendered image and the scene semantic field, and transfer the style of the scribbled image to 3D scene with the obtained semantic correspondence.

For that, we compose a modified style dictionary $\mathcal{D}$ with the clustered semantic features ${\mathbf{f}}_\text{DINO}(\textbf{I}^{j}_\text{ren})$ extracted from the rendered image $\textbf{I}_\text{ren}$ and the style codes from the scribbled image $\textbf{I}_\text{scr}$, as \(\mathcal{D} {=} \{ \bar{\mathbf{f}}_\text{DINO}(\mathbf{I}_\text{ren}^{j}){:}(\mu_\text{scr}^{j}, \sigma_\text{scr}^{j})\}_{j=1,\dots,M} \).
A centroid of DINO feature $\bar{\mathbf{f}}_\text{DINO}(\textbf{I}^{j}_\text{ren})$ is 
averaged from each cluster. 
A style code $(\mu_\text{scr}^{j}, \sigma_\text{scr}^{j})$ is a pair of the mean and standard deviation of the VGG features extracted from each cluster of $\textbf{I}_\text{scr}$, which have the same clustered region with $\textbf{I}_\text{ren}$.
Then we compute semantic correspondence and a semantic-weighted style code using Eq. (13) with the modified dictionary $\mathcal{D}$.
As a result, we obtain the semantic-weighted style code (\(\textbf{M}_w\), \(\mathbf{\Sigma}_w\))
being assigned to 3D Gaussians, which 
contains semantically corresponding style codes from the scribbled image $\mathbf{I}_\text{scr}$.
Then we can obtain the stylized result using Eq. (14), which transfer the style of scribbled image to the scene, by considering semantic correspondence between the rendered image and the 3D scene. 

\begin{figure*}[t]
    \centering
    \includegraphics[width=\linewidth]{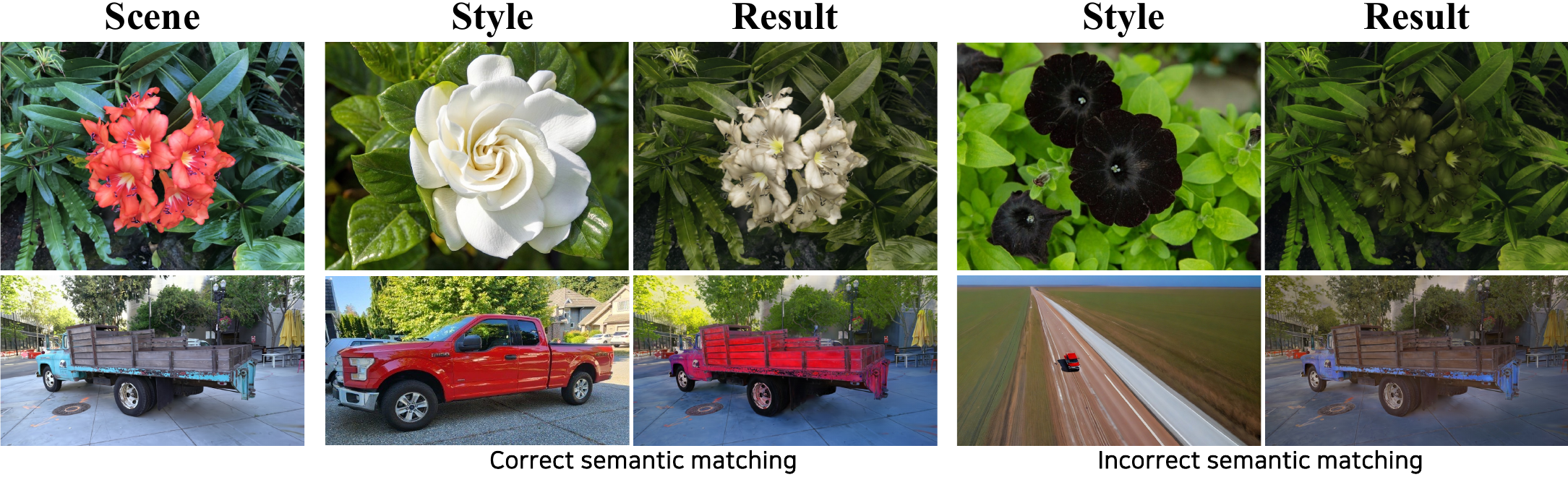}
    \caption{\textbf{Stylization results with correct semantic matchings and incorrect semantic matchings.} \textbf{[Left]} Original 3D scene. \textbf{[Mid]} Stylization results with correct semantic matchings. \textbf{[Right]}
    Failure cases of our methods with incorrect semantic matchings in extreme cases.
    }
    \label{fig:supp_failure_cases}
\end{figure*}

\section*{Appendix B \\ Implementation details}
\paragraph{Generalizable pre-trained MLP color decoder}
We propose a pre-trained MLP color decoder which enables feed-forward style transfer with new arbitrary styles (see Sec. \uppercase\expandafter{\romannumeral4}-A).
We compose the decoder with a sequence of 2 linear layers whose output channels are 128 and 3, followed by a sigmoid function to output an RGB value.
The color decoder is trained on the MS COCO~\citep{lin2014microsoft} dataset and the wikiart~\citep{nichol2016painter} dataset as a content and a style image dataset.
We train the network for 50,000 iterations with the learning rate 0.0001.
\begin{figure}[thbp]
    \centering
    \centering
        \includegraphics[width=\linewidth]{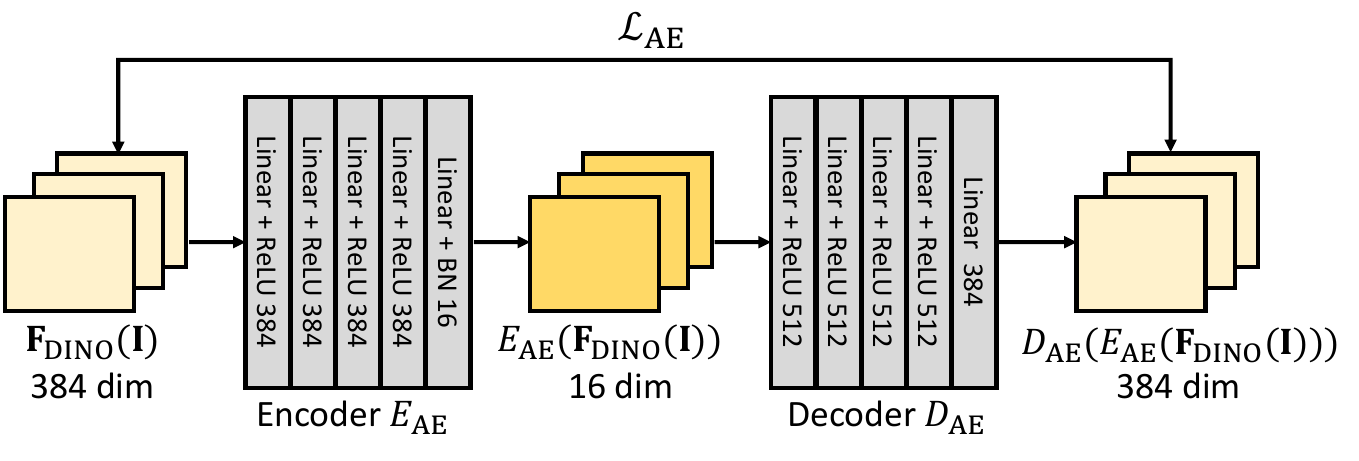}
        \caption{   \textbf{Training pipeline of the semantic feature autoencoder.} $\textbf{F}_\text{DINO}(\textbf{I})$ denotes the feature maps extracted from the training images $\textbf{I}$ with DINO and $\mathcal{L_\text{AE}}$ denotes the reconstruction loss (see Eq. (12)).
        }
\label{fig:supp_autoencoder}

\end{figure}

\paragraph{Semantic feature autoencoder}
We employ a scene-specific semantic feature autoencoder compressing high-dimensional semantic features to low-dimensional features, which enables semantic feature field represented by 3D Gaussians (see Sec. \uppercase\expandafter{\romannumeral5}).
For the DINO~\citep{caron2021emerging} features whose channel size is 384, we set the channel of low-dimensional features to 16. 
Figure \ref{fig:supp_autoencoder} shows the detailed architecture and the training process of the semantic feature autoencoder.
We train the autoencoder for 5 epochs with semantic features extracted from training images of each scene.
We set the learning rate to 0.0001.

\section*{Appendix C \\ Additional results}
\paragraph{Failure cases}
As mentioned in Sec. \uppercase\expandafter{\romannumeral7}, our semantic matching performance is bounded by the semantic image encoder, DINO~\citep{caron2021emerging}. Figure~\ref{fig:supp_failure_cases} shows failure cases of our method, which are related to the capability of DINO. The first row shows a success case and a failure case due to an out of distribution sample, a black flower. DINO cannot correctly predict that the black flower in the reference image and the red flower in the original scene are in the same category. \ie, have semantic correspondence. 
The second row also shows that DINO works well when the truck in the reference image is big enough, however, fails to match the extremely small truck on the road and the truck in the original scene.

\begin{figure*}[thbp]
    \centering
        \includegraphics[width=\linewidth]{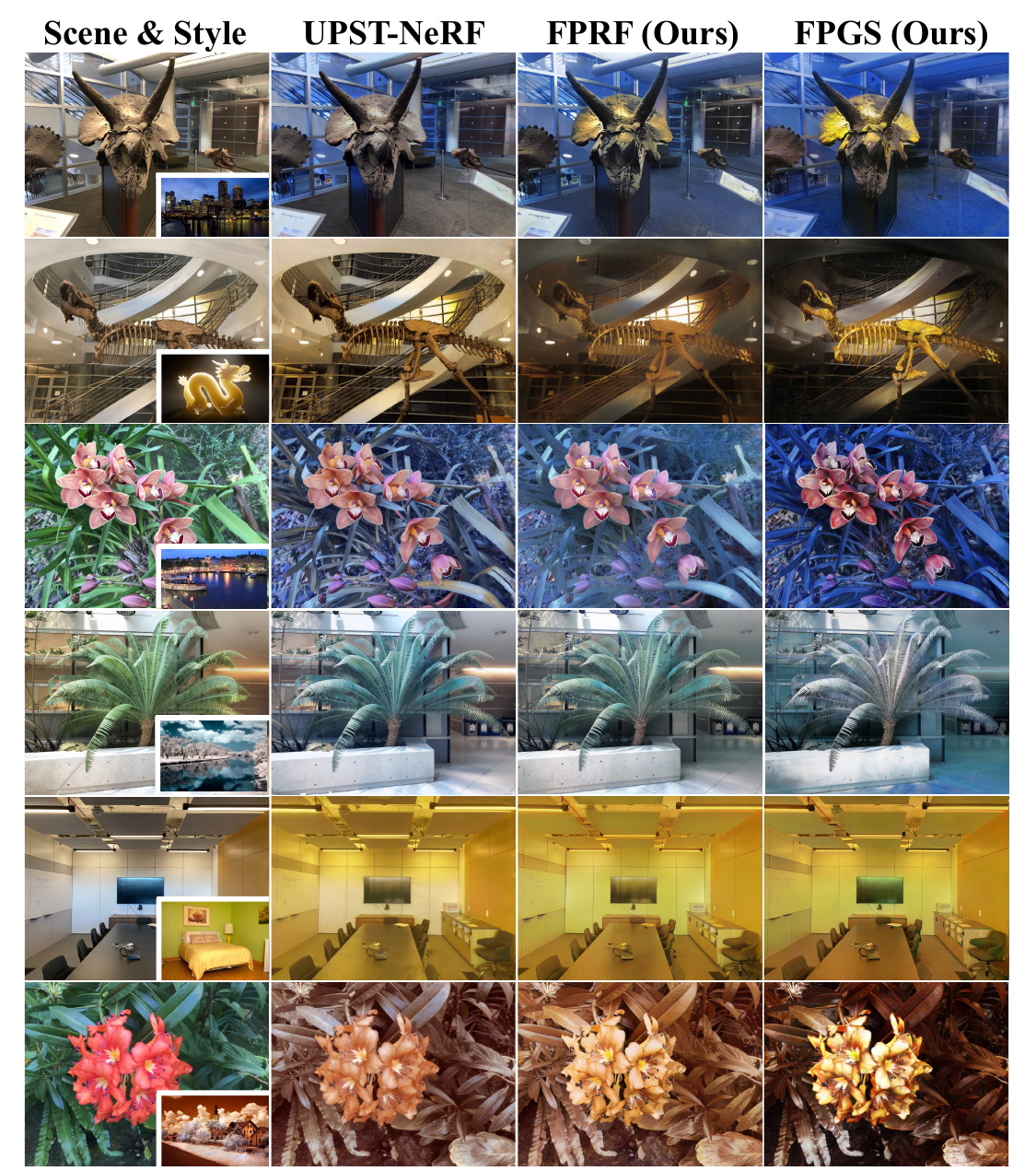}
                \caption{\textbf{Additional qualitative results on LLFF dataset~\citep{mildenhall2019local}.}
        Compared to the UPST-NeRF~\citep{chen2024upst}, our methods accurately reflect the diverse color of the reference image.}
        
\label{fig:supp_LLFF}
\end{figure*}

\begin{figure*}[thbp]
    \centering
    \centering
        \includegraphics[width=\linewidth]{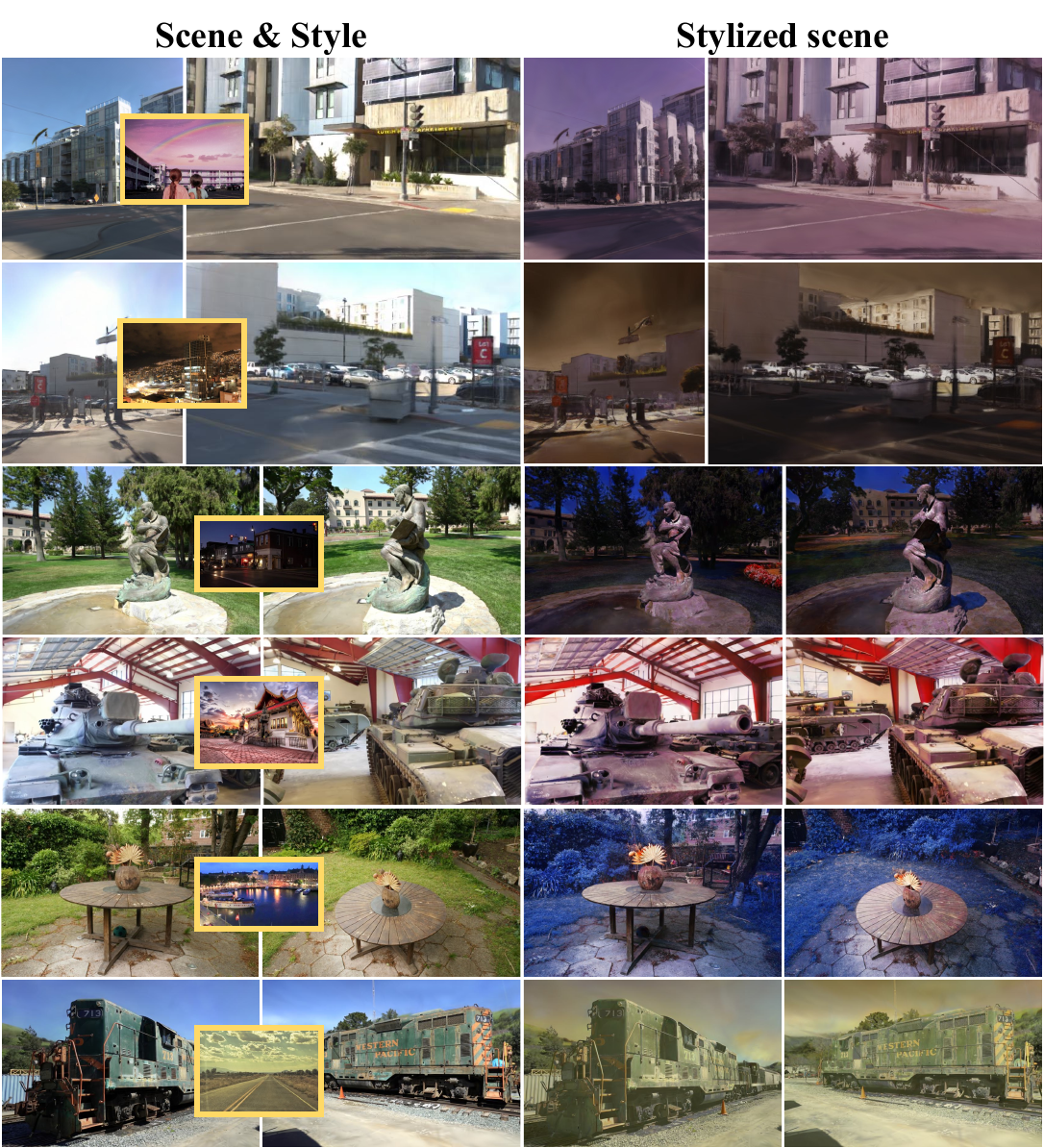}
        \caption{   \textbf{Additional qualitative results on unbounded scenes.}
           Style transfer results on the San Francisco Mission Bay dataset~\citep{tancik2022blocknerf} (top 2), on the Tank and Temples dataset~\citep{knapitsch2017tanks} (mid 2), and on the Mip-NeRF 360 dataset~\citep{barron2022mip} (bottom 2).
        }
    \vspace{3mm}
\label{fig:supp_large}

\end{figure*}

\begin{figure*}[thbp]
    \centering
    \centering
        \includegraphics[width=\linewidth]{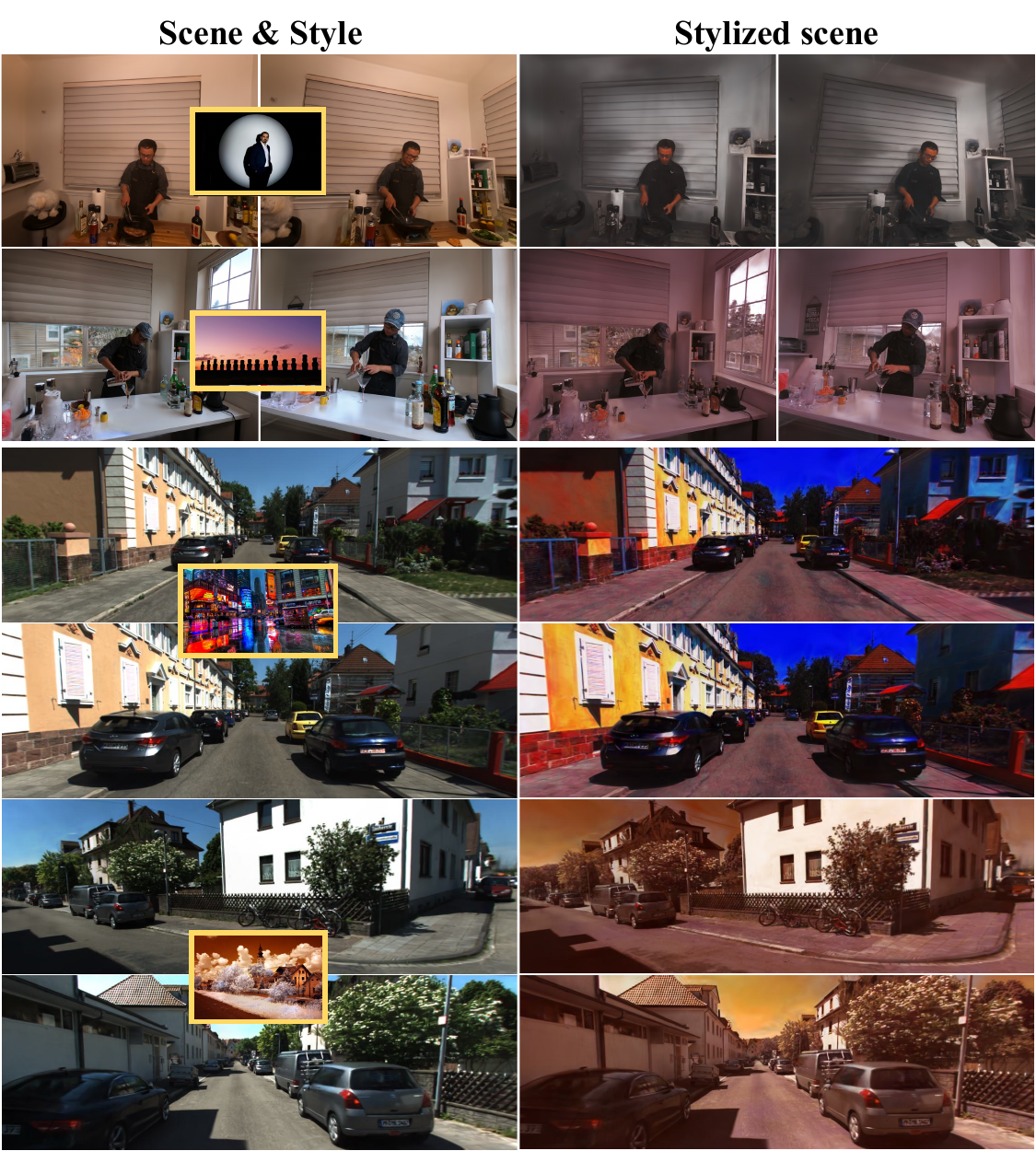}
        \caption{   \textbf{Additional qualitative results on 4D scenes}
           Style transfer results on the DyNeRF dataset~\citep{li2022neural} (top) and on the KITTI-360 dataset~\citep{liao2022kitti} (bottom).  
        }
\label{fig:supp_4d}

\end{figure*}

\paragraph{Qualitative results} Figure~\ref{fig:supp_LLFF} shows comparisons with UPST-NeRF~\citep{chen2024upst} on the LLFF dataset~\citep{mildenhall2019local}.
Figure~\ref{fig:supp_large} shows qualitative results on the 
San Fran Cisco Mission Bay dataset~\citep{tancik2022blocknerf}, the Tank and Temples dataset~\citep{knapitsch2017tanks}, and the Mip-NeRF 360 dataset~\citep{barron2022mip}, which contain unbounded scenes.
Figure~\ref{fig:supp_4d} shows qualitative results on the DyNeRF dataset~\citep{li2022neural} and the KITTI-360 dataset~\citep{liao2022kitti} dataset, which are 4D scene datasets.


\bibliographystyle{spbasic}
\bibliography{supplement}